\documentclass[aps, reprint]{revtex4-2}
\usepackage{amsmath,amsbsy,amssymb,amsfonts}
\usepackage{graphicx, color}
\usepackage{empheq}
\usepackage{float}
\usepackage{hyperref}
\usepackage[capitalise]{cleveref}
\usepackage{subfigure}
\usepackage{natbib}
\usepackage{babel}

\newcommand{\Matr}[1]{\boldsymbol{\mathcal{\hat{#1}}}}
\newcommand{\angstrom}{\text{\normalfont\AA}}
\newcommand{\sech}{\text{\normalfont sech}}
\newcommand{\crn}[1]{\hat{#1}^{\dagger}}
\newcommand{\anh}[1]{\hat{#1}^{}}

\newcommand{\prtl}{\partial}

\graphicspath{{figures/}}

\begin{document}
\newcounter{tablerow}
\crefname{tablerow}{row}{rows}
\date{\today} 

\title{The mechanism of electrical conduction in glassy
  semiconductors}

\author{Arkady Kurnosov}\affiliation{Department of
  Chemistry, University of Houston, Houston, TX 77204-5003}  \affiliation{Department of Physics, Wesleyan University, Middletown, CT 06459}

\author{Vassiliy Lubchenko} \email{vas@uh.edu}
\affiliation{Department of Chemistry, University of Houston, Houston, TX 77204-5003} \affiliation{Department of Physics, University of Houston, Houston, TX 77204-5005}
  \affiliation{Texas Center for Superconductivity, University of Houston, Houston, TX 77204-5002}


\begin{abstract}

We argue that the dominant charge carrier in glassy semiconducting alloys is a compound particle in the form of an electron or hole bound to an intimate pair of topological lattice defects; the particle is similar to the polaron solution of the  Su-Schrieffer-Heeger Hamiltonian.  The spatial component of the density of states for these special polarons is determined by the length scale of spatial modulation of electronegativity caused by a separate set of standalone topological defects. The latter length scale is fixed by the cooperativity size for structural relaxation; the size is largely independent of temperature in the glass but above melting, it decreases with temperature. Thus we predict that the temperature dependence of the electrical conductivity should exhibit a jump in the slope near the glass  transition; the size of the jump is predicted to increase with the fragility of the melt. The predicted values of the jump and of the conductivity itself are consistent with experiment.
\end{abstract}

\maketitle
The microscopic mechanism of electrical conduction in glassy semiconductors is a long-standing question of condensed matter physics and materials science. Here we focus on the important class of glassy semiconductors exemplified by the chalcogenide alloys and similar inter-metallic compounds. Optoelectronic  properties of these fascinating compounds are of great interest both mechanistically~\cite{PWA_negU, KAF, PhysRevB.23.2596, RevModPhys.50.209, Emin_rev, Emin_revII, EminSwitching, ZL_JCP, ZLMicro1, ZLMicro2, GHL, LL1, LL2, Konstantinos2019, manjon2024electron, lee2024bonding, raty2024tailoring} and in applications: Best known as phase-change memory (PCM) materials, they are a candidate system for the next generation non-volatile solid-state memory~\cite{PhysRevLett.21.1450, Kolobov2004, Steimer2008, WuPop2024, https://doi.org/10.1002/appl.202200024, Yarema2024}, re-writable data storage~\cite{ISI:000250615400019}, parallel neuromorphic computing architectures~\cite{AIST, https://doi.org/10.1002/advs.202406433, WOS:001045744300001}, ultrafast solid-state displays, semi-transparent smart glasses, smart contact lenses, and artificial retina devices~\cite{Hosseini2014} among others~\cite{PRABHATHAN2023107946}. In reflection of their technological importance, the market for the PCM materials is projected to continue growing at 25\% or greater, annually, through year 2033~\cite{PCMmarket2024}, consistent with the growing numbers of related patents~\cite{PCMpatents2024}.

In Mott's picture \cite{MottDavis1979, Mott1993}, electrical current in amorphous semiconductors is carried through relatively extended, wavepacket-like electronic excitations. The lattice remains largely a spectator of electronic motions, vibrations being a perturbation as in Migdal's theorem~\cite{ISI:A1958WX85700010}. Owing to the disorder, the electronic bands are not collections of infinitely-extended Bloch states, but, instead, are {\em mobility} bands composed of orbitals whose extent only needs to be greater than the mean free path of the electronic quasiparticle.  Emin~\cite{Emin_rev, Emin_revII} argued that when the electron-lattice coupling is sufficiently strong, the current is carried, instead, by small polarons~\cite{PhysRevLett.36.323}. The small polaron is a compound, emergent entity in the form of a self-trapped electron or hole that occupies an impurity-like bound state. The bound state is due to a local deformation of the lattice stabilized by the trapped charge itself; its formation is subject to an activation barrier~\cite{PhysRevLett.36.323}. Mott's and Emin's scenarios each predict an Arrhenius temperature dependence of the conductivity, the activation energy tied to the optical gap. Both descriptions are continuum; the characteristic length scale for the density of states is supplied by the spatial concentration of the frontier orbitals, irrespective of the material's structure.  

The optical gap~\cite{ARAI1975295, Hosokawa_1991, LKgap} and the structure~\cite{LL1, LW_aging} of a glassy melt both vary continuously with temperature across the glass transition. The treatments in Refs.~\cite{Mott1993, Emin_rev} thus imply the electrical conductivity should depend smoothly on temperature near the glass transition temperature $T_g$. Contrary to this expectation, the {\em measured} temperature dependence of the electrical conductivity $\sigma$ exhibits a distinct jump in the slope at $T_g$~\cite{SeagerQuinn}, see Fig.~\ref{sigmaT}  and Figs.~S3 and S4 in the Supplementary Information (SI).

\section*{Microscopic picture}

To rationalize this apparent, puzzling connection between electronic and structural properties of glassy semiconductors, here we put forth an expressly non-continuum 

\hspace{-5.5mm} picture. We set the stage by first reviewing several relatively recent findings the present picture builds upon.

Glassy melts and frozen glasses alike exhibit organization on two length scales. The smaller of these length scales, often called the bead size $a$~\cite{XW, LW_soft},  corresponds with the first sharp diffraction peak~\cite{ISI:A1994NL31200006, ElliottNature1991} and is only marginally greater than the atom spacing. This length is static and reflects a symmetry breaking at the level of the first coordination shell~\cite{LL1} caused by the anisotropy in bonding intrinsic to electron rich centers~\cite{PapoianHoffmann2000, GHL, ABW}. The pertinent particle spacing in the symmetry-lowered structure can be thought of as the size of a rigid molecular unit that is perturbed only weakly during structural relaxation~\cite{LL1, BL_6Spin}.  The greater length, often called the cooperativity size $\xi$~\cite{LW}, is the volumetric size of the smallest region than can reconfigure in the glassy material~\cite{KTW, LW_aging, LRactivated}.  In a frozen glass, $\xi \simeq 6a \simeq 2 - 3$~nm, $1/\xi^3 \simeq 10^{20}$~cm$^{-3}$~\cite{RWLbarrier}, approximately independent of temperature~\cite{LW_aging}. Upon melting, the length $\xi$ begins to decrease  with temperature and drops down to $3a$ or so, in magnitude, by the dynamical crossover, beyond which the system becomes a uniform liquid~\cite{LW_soft, SSW, RL_LJ}. This microscopic picture has quantitatively explained dozens of disparate phenomena in glassy melts and frozen glasses, see reviews~\cite{LW_ARPC, L_AP, LW_RMP, Lrelics}. Though of dynamical origin~\cite{LW_ARPC, XWbeta, LWphoto}, the length $\xi$ has a static aspect in that it provides the characteristic length scale for the spatial variation of the built-in strain in the glassy matrix~\cite{LRactivated, L_AP}. 

\begin{figure}[t]
\centering
  \includegraphics[width= .95 \linewidth]{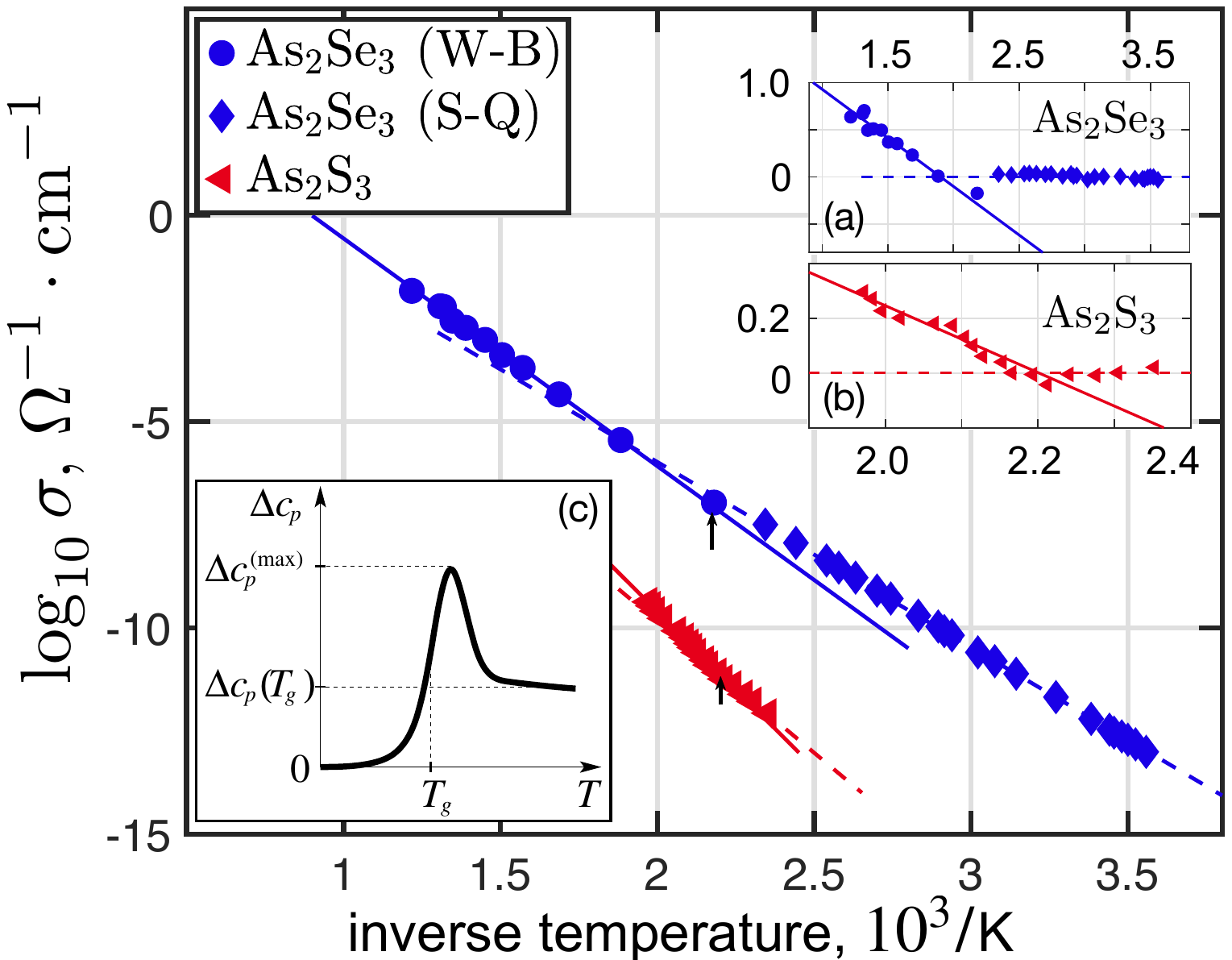}
  \caption{Electrical conductivity as a function of the inverse temperature for select chalcogenide alloys: As$_2$Se$_3$~\cite{WebbBaker1972,SeagerQuinn} and As$_2$S$_3$\cite{Bobb1975techreport}. We used $T_g$ values 460~K and 465~K~\cite{Davey1983, doi:10.1111/j.1151-2916.1977.tb15488.x, https://doi.org/10.1002/pssb.19640070302}, respectively, shown by arrows, to obtain the fits; see SI for details. In insets (a) and (b), the respective slopes of the $T<T_g$ segments are subtracted for clarity. Inset (c): A generic sketch of the $T$-dependence of the excess heat capacity of a melt, relative to the glass, for an upward temperature scan.}
  \label{sigmaT}
  \vspace{-3mm}
\end{figure}

\begin{figure}[t]
  \includegraphics[width= 0.9 \linewidth]{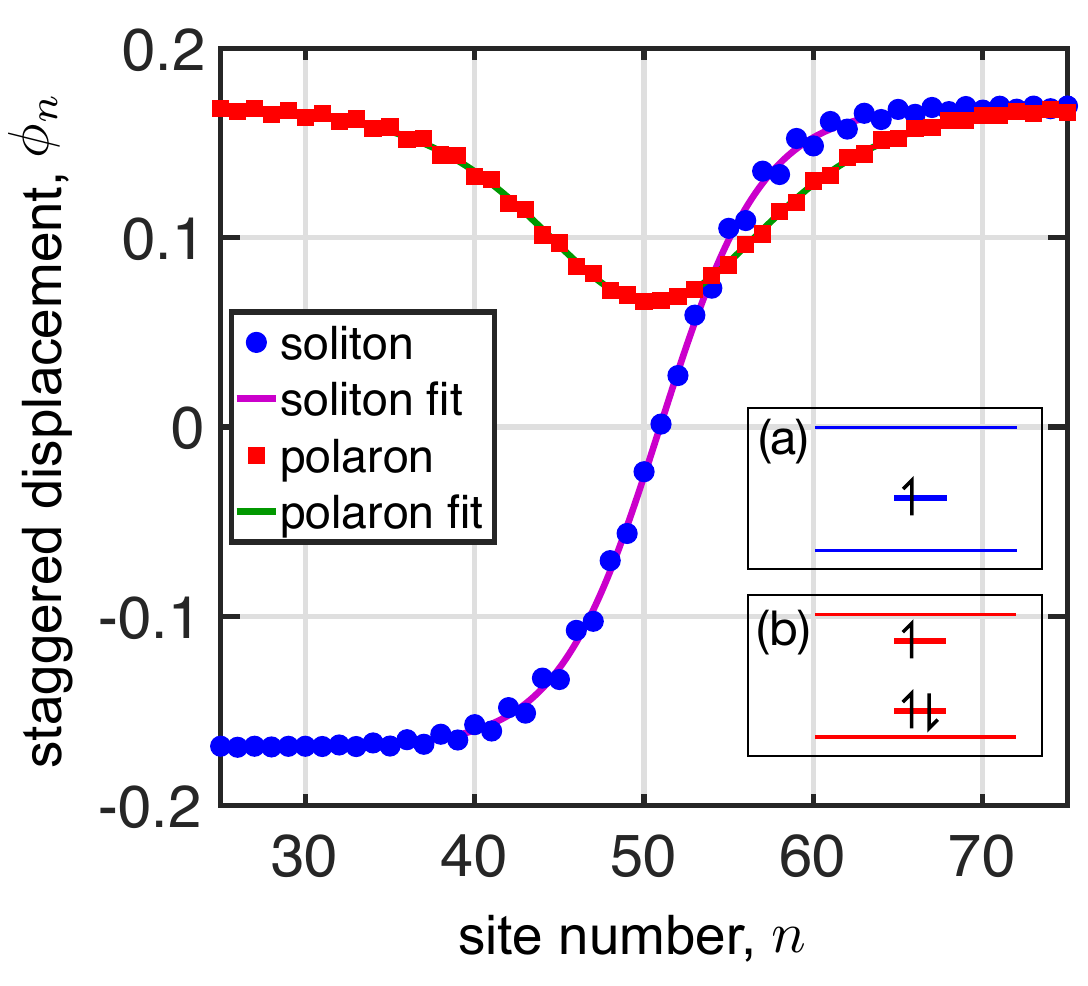}
  \caption{Two numerical solutions of the SSH Hamiltonian~\cite{RevModPhys.60.781}: The staggered displacement $\phi_n \equiv (-1)^n u_{n}$ as a function of site $n$, where $u_n$ is the displacement of site $n$ off its value in the Peierls-unstable, vibrational ground state. Circles: an odd-numbered chain at half filling. Squares: an even-numbered chain with an extra electron. Solid lines are functional fits. The corresponding electronic midgap states are shown in insets (a) and (b), respectively. The parameters of the Hamiltonian were chosen so as to approximately match the insulating gap in arsenic sulfide and selenide, see Supplemental Information (SI).}
  \label{polaron}
  \vspace{-3mm}
\end{figure}

\begin{figure}[t]
  \includegraphics[width= \linewidth]{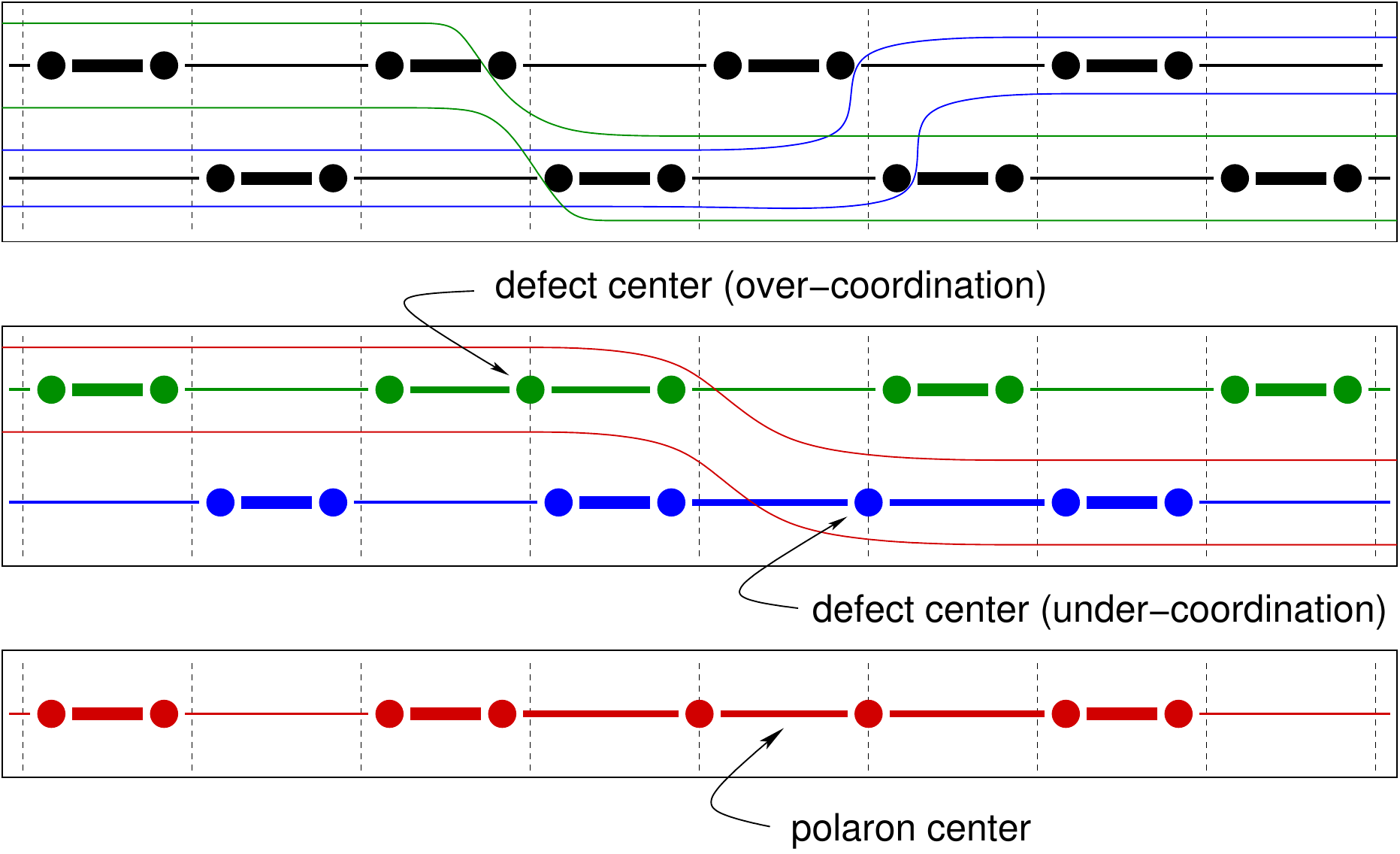}
  \caption{A simplified schematic of particle displacements and bond strengths to accompany Fig.~\ref{polaron}. Atoms are denoted with circles, bonds with horizontal straight lines connecting the atoms. Of the three bond types shown, the thickest line corresponds to the covalent bond, the thinnest line to a closed-shell secondary interaction, and the intermediate thickness corresponds to an intermediate bond strength. {\em Top panel:} The two chains correspond to the two alternative ways to form a perfectly dimerized ground state configuration. {\em Middle panel}: Two distinct ways to form a solitonic defect---color-coded green and blue---by interfacing the two ground state configurations from the top panel in two different ways, so as to form an over-coordinated and under-coordinated center, respectively. {\em Bottom panel:} An polaron configuration can be thought of as resulting from either a pair of two solitonic defects or from stretching a strong bond in a perfectly dimerized chain, and then adding an electron or hole.} 
  \label{SSHexpln} 
  \vspace{-3mm}
\end{figure}

We have argued that in glassy semiconductors that are charge-density-wave (CDW) solids, such strained regions host midgap electronic states~\cite{ZL_JCP, ZLMicro2, LL2}. These special midgap states are similar to the topological midgap states arising when the two alternative dimerization patterns of a trans-polyacetylene chain are brought into contact~\cite{ZL_JCP, RevModPhys.60.781}. 
This picture provides a unified, quantitative explanation for the puzzling light-induced midgap absorption and electron paramagnetic resonance signal~\cite{BiegelsenStreet, PhysRevB.38.11048}, anomalous fluorescence~\cite{TadaNinomiya, TadaNinomiya2, TadaNinomiya3}, and difficulty in doping glassy semiconductors~\cite{PWA_negU, Kolomiets1981}. We illustrate here the emergence of these special midgap states by considering a Peierls-distorted chain of electron-hosting sites at half-filling, at the level of the Su-Schrieffer-Heeger  (SSH) Hamiltonian~\cite{RevModPhys.60.781}. The spatial profile of the staggered displacement of the sites, in the ground state of an odd-numbered, closed chain is shown with circles in Fig.~\ref{polaron}; see also an explanatory schematic in Fig.~\ref{SSHexpln}. There is a topologically stable defect in the dimerization pattern---owing to the odd number of sites---that cannot be removed by elastic deformation. The conduction and valence edges of the insulating gap effectively cross at the defect, thus leading to the appearance of a midgap electronic state there~\cite{PhysRevD.13.3398, RevModPhys.60.781, LL2}. Inset (a) of Fig.~\ref{polaron} shows a neutral midgap state with spin $1/2$; it can be thought of as a solid-state analog of a free radical. 
  
 A topological midgap state in a glassy semiconductor is typically charged negatively or positively~\cite{ZL_JCP, ZLMicro2, LL2}, which would correspond to the singly-occupied state in Fig.~\ref{polaron}, inset (a), becoming instead either filled or empty, respectively. The so charged defects thus comprise a disordered checkerboard pattern of excess negative and positive charges, the corresponding length scale being the cooperativity length $\xi$ itself. The pattern is tied to the structure and can change only if the melt flows or ages; both of the latter processes are activated and, typically, slow~\cite{LW_ARPC, L_AP}. This charge distribution is inherently built-in owing to the structural degeneracy of the disordered lattice; the emergence of this charge distribution cannot be captured by continuum treatments~\cite{XW, L_AP}.
 
Now, the present picture advances the following  microscopic notions, to be detailed below: On the one hand, the charge carriers are topological polarons each in the form of an electron or hole bound to an intimate pair of topological defects. On the other hand, the inherent aperiodic charge pattern is a source of an electrostatic potential, see Fig.~\ref{pot}(a), that will act on itinerant charges. Furthermore, the variation of the aperiodic electric field is sufficiently large to cause transient localization of the polarons; the characteristic length scale of spatial variation of the field, then, determines the density of states for charge carriers.

To establish the detailed nature of itinerant charges in glassy CDW solids, we note that already their periodic counterparts can house intimate pairs of topological defects bound to a polaron~\cite{Su5626, ISI:A1981MD41000002, RevModPhys.60.781}. We illustrate such an (electron) polaron configuration in Fig.~\ref{polaron}, at the SSH level, using an even-numbered closed chain at half-filling and an added electron. In the extreme limit of a chain of weakly interacting dimers, this special polaron can be thought of as an electron placed in the anti-bonding orbital on an individual dimer, inset (b) of Fig.~\ref{polaron}. At the same time, a destabilization of the filled orbital takes place. The bonding---anti-bonding orbital pair corresponds to two mutually-hybridized midgap states centered on, respectively, an over-coordinated and under-coordinated site~\cite{RevModPhys.60.781}. The {\em hole}-polaron case is analogous and would correspond to an empty antibonding orbital and half-filled bonding orbital in Fig.~\ref{polaron}, inset (b). 

An extended carrier turns into the topological polaron via a downhill, barrierless process that results in a relatively extended deformation pattern, shown with squares in Fig.~\ref{polaron}. {Indeed, this pattern can be thought of as obtained by stretching one strong bond in a perfectly dimerized chain while adding an electron or hole to the bond and adjusting the rest of the chain accordingly; c.f. the top and bottom panel, respectively, of Fig.~\ref{SSHexpln}.  Polaron-like configurations analogous to that in Fig.~\ref{polaron} have been obtained at an {\em ab initio} level in 3D samples of disordered chalcogenide alloys~\cite{LL2}. The corresponding orbitals are part of the Urbach-Lifshitz tail of the localized states~\cite{LL2, LKgap} and can contain one or more quasi-linear fragments, each of which is similar to the topological polaron in conjugated polymers~\cite{Su5626, ISI:A1981MD41000002, RevModPhys.60.781}.  

The electrical conductivity is a sum of the contributions from the electron ($e$) and hole ($h$) polarons, respectively:
\begin{equation} \label{Eq:cond}
  \sigma = \int \frac{dE}{k_B T} [q_e n_e(E) \mu_e(E) + q_h n_h(E)  \mu_h(E) ]
\end{equation}
where $n_i(E)$ is the density of thermally available polaron states, $q_i$ the effective charge, and $\mu_i(E)$ the mobility, respectively, of carrier $i$ at energy $E$.

\begin{figure}[t]
  \includegraphics[width=\linewidth]{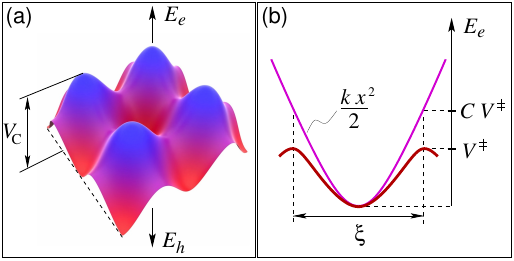}
  \caption{(a) 2D caricature of the electrostatic potential due to the intrinsic, charged defects. $E_e$ and $E_h$ denote the energy of an electron and hole, respectively. (b) Graphical explanation of the derivation of Eq.~(\ref{Vddagger}).}
  \label{pot} 
\end{figure}

\section*{The polaron density of states depends on temperature}

The polaron must be centered on a covalent bond, per
Fig.~\ref{polaron}, inset (b). Each pnictogen typically forms three such bonds, chalcogen two~\cite{ZLMicro1}. The bonding is locally distorted-octahedral, where the number of covalent bonds is approximately a half of the total number of the inter-atomic contacts, the rest being weaker, closed-shell interactions~\cite{ZLMicro1, LL1}. At the same time, the covalent bonds must be associated with rigid molecular units; thus their volumetric volume should be equated with the bead size $a$; generically $a \simeq 4\!$~\AA ~in chalcogenides~\cite{ZL_JCP}. Pretend for a moment that the underlying lattice and the potential due to the built-in charged topological defects, Fig.~\ref{pot}(a), are both strictly periodic and have the same point-group symmetry. The lowering of the translational symmetry, due to the potential, would result in a reduction of the Brillouin zone, relative to the potential-free case, by a factor of $(\xi/a)$ along each spatial dimension. The new, smaller zone would then contain $(\xi/a)^3$ distinct polaron bands per each band in the original, large zone. 

The potential exhibits enough variation, $V_\text{C} \sim e^2/\xi \gtrsim 0.1$~eV, for the polaron (mobility) bands to become so narrow as to break up into distinct sets of vibrational levels within respective wells. (This is similar to how core electronic orbitals in a solid do not form bands but, instead, remain localized.) That the potential is aperiodic will only enhance the localization. In setting up a systematic argument, we first note that the potential around an individual charged defect is parabolic in the vicinity of its respective minimum, denote the corresponding spring constant with $k$. Thus the vibrational quantum number $n$ for motion in {\em one} spatial direction is limited from above, $n \leqslant n_\text{max}$ so that $\hbar \omega_s n_\text{max} = C V^\ddagger$, where $V^\ddagger$ is the altitude of the lowest saddle point relative to the bottom of the well, $\omega_s = (k/m_s)^{1/2}$, $m_s$ the soliton's mass, and $C \approx 2$ is a numerical constant. The latter constant specifies, by construction, how much the parabolic approximation exceeds the actual potential near the saddle point: $k (\xi/2)^2/2 \equiv C V^\ddagger$, see Fig.~\ref{pot}(b). The total number of the vibrational states for {\em three}-dimensional motion is approximately $n_\text{max}^3/6$. To estimate the value of $V^\ddagger$ above which every polaron band must cross classically forbidden regions, we require that the number of bands $(\xi/a)^3$ be less than $n_\text{max}^3/6$.  The mass of the soliton is significantly lower than the atomic mass $m_a$, because a displacement of the soliton over a lattice spacing implies a much smaller displacement, $d_L$ or so, for an actual atom~\cite{LW, RevModPhys.60.781}: $m_s = (d_L/a)^2 m_a$. The length $d_L$, often called the Lindemann length~\cite{L_Lindemann}, is the displacement at the mechanical stability edge of a solid. We set $C=2$, for concreteness, to obtain:
\begin{equation} \label{Vddagger} V^\ddagger \sim 30 \frac{\hbar^2}{2 m_a d_L^2}.
\end{equation}
Using $a = 4\, \angstrom$ and $m_a/m_e = 10^5$, where $m_e$ is the electron mass, one obtains $V^\ddagger \sim 7 \cdot 10^{-3}$~eV, which is much less than $V_c$. As a consequence, the tunneling matrix element for inter-well tunneling is astronomically small: $\sim \exp[-(2 m_s V_\text{C})^{1/2} \xi/\hbar] \sim \exp[ -(V_\text{C}/V^\ddagger)^{1/2} (3 \xi/a) ] \sim 10^{-30}$, where we used the standard WKB expression for the under-barrier wavefunction and Eq.~(\ref{Vddagger}).

The localization of the polarons means that the translational degrees of freedom of the itinerant charge have been effectively converted into vibrations of the polaron around an individual charged center. The expectation value of the polaron's vibrational coordinate is at the center of the respective well of the potential from Fig.~\ref{pot}(a), be the vibration in the ground or any of the excited states. As a result, the location of the polaron can be only determined up to the spacing between nearby centers, i.e., the length $\xi$ itself. Thus the spatial component of the density of states for charge carriers must be equal to $1/\xi^3$, not the usually postulated density of valence electrons/holes. 

At the same time, because the tunneling between distinct minima is negligible, the polaron must move via thermally-activated, adiabatic transitions~\cite{FW} between nearby minima on the potential surface from Fig.~\ref{pot}(a). An individual jump is thus of length $\xi$. At the transition state, the electron/hole is delocalized at least over a distance $\xi$ and is best thought of as belonging to the pertinent mobility band. The activation barrier for polaron's hopping between pairs of distinct attractive wells is then determined by the distance from the polaron level ($E_e$ for electrons, $E_h$ for holes) to the edge of the appropriate mobility band ($E_c$ conduction, $E_v$ valence band). Thus we obtain for the electron (hole) polaron hopping rate: $\tau^{-1} = \tau_0^{-1} e^{-\beta (E_c-E_e)}$ ($\tau^{-1} = \tau_0^{-1} e^{-\beta  (E_h-E_v)}$), and for the (long-time) diffusivity $D = \xi^2/\tau$. To determine the prefactor $\tau_0$ we note that the bonds must return to their equilibrium lengths when the polaron delocalizes. Thus the reaction coordinate near the transition state is essentially the vibration of the covalent bond where the polaron is centered at the moment, denote the respective frequency with $\omega_B$. Away from the transition state, the reaction coordinate increasingly hybridizes with other vibrational modes, implying a barrier-crossing event will have likely succeeded after a single vibration of the reaction coordinate. Consequently the prefactor can be well approximated by its value near critical damping, $\tau_0^{-1} = \omega_B/2\pi$~\cite{FW}. The energy dependence of the equilibrium density of states $n(E)$ for electrons, at the energies in question, is given by the Boltzmann factors $e^{-\beta (E-\mu)}$, $\mu$ being the chemical potential. The overall activation rate determining the conductivity in Eq.~(\ref{Eq:cond}) becomes the Arrhenius factor $e^{-\beta (E_c-\mu)}$ for electrons and $e^{-\beta (\mu - E_v)}$ for holes. Thus Eq.~(\ref{Eq:cond}) yields:
\begin{equation} \label{Eq:cond2}
  \sigma = \frac{q^2_e \omega_B}{2 \pi \xi k_B T} \left[e^{-\beta
      (E_c-\mu)} + e^{-\beta (E_v-\mu)} \right].
\end{equation}
In most cases, the lack of electron/hole symmetry implies the conduction is dominated by either the electron or hole polarons; for instance in arsenic chalcogenides, the hole is the main carrier~\cite{Emin_Hall}. For this reason, only the dominant term inside the square brackets should be kept.

\section*{Comparison with available experiments}

In discussing implications of Eq.~(\ref{Eq:cond2}) for the $T$-dependence of $\sigma$, we begin with an idealized experimental protocol, in which the temperature is scanned across the glass transition while little aging takes place. According to the random first-order transition (RFOT) theory~\cite{L_AP, LW_ARPC}, the cooperativity size $\xi$ will remain near stationary in the glass but will be decreasing with temperature above melting, a good approximation provided by the expression~\cite{LRactivated} $\xi = (K k_B/ T s_c^2)^{1/3}$. Here $s_c$ is the configurational entropy per unit volume and $K$ is the bulk modulus. As a result, the $T$-derivative of the length $\xi$ and, hence, of the conductivity $\sigma$ should exhibit a discontinuity at the glass transition. The discontinuity is somewhat smeared because the relaxation times in the glass are distributed~\cite{XWbeta, Lionic}. Thus we obtain a simple relation $\xi(T) = \xi(T_g) [s_c^2(T_g) T_g/K(T_g)]^{1/3} [s_c^2(T) T/K(T)]^{-1/3}$ for $T \geqslant T_g$ and $\xi(T) = \xi(T_g)$ for $T < T_g$. This yields that the apparent activation energy $E^\ddagger = - \prtl \ln \sigma(T)/\prtl (1/k_B T)$ should exhibit a discontinuity $\Delta E^\ddagger \equiv E^\ddagger(T_g^+)-E^\ddagger(T_g^-)$ at the glass transition:
\begin{equation}\label{Edjump}
\Delta E^\ddagger  = \left[2 \frac{ \Delta c_p^\text{(max)}}{s_c(T_g)} + 1 - \left( \frac{\partial \ln K}{\partial \ln T} \right)_{T_g^+} \right] \frac{k_B T_g}{3}.
\end{equation}
Here $\Delta c_p^\text{(max)}$ is the peak value of the excess heat capacity $\Delta c_p(T) = T (\partial s_c/\partial T)$ of the liquid relative to the frozen glass. Conductivity data are often collected using aged samples, however, which informs the following discussion. 

During melting of a frozen glass, $\Delta c_p(T)$ undergoes a (smeared) jump from zero to a positive value and, then, gradually declines with temperature. The latter gradual decline is often preceded by a sharp peak, see illustration in Fig.~\ref{sigmaT}, inset (c). This peak-like overshoot, if any, comes about because glasses are away-from-equilibrium systems that relax toward the lowest free-energy state pertinent to the ambient temperature~\cite{LW_aging, SWultimateFate, L_AP}. The height $\Delta c_p^\text{(max)}$ of the peak is the greater, the more slowly the glass had been prepared and/or the longer the prepared glass had been allowed to age~\cite{doi:10.1111/j.1151-2916.1977.tb15488.x}. By construction,
\begin{equation} \label{cpineq}
    \Delta c_p^\text{(max)} \geqslant \Delta c_p(T_g),
\end{equation}
the equality pertaining to a freshly made glass that was made relatively quickly, see a concrete example in SI.  Eq.~(\ref{Edjump}) thus indicates that the jump in the apparent activation energy $E^\ddagger$ will exhibit a range of values depending on the extent of aging, samples undergone more aging corresponding to greater values of $E^\ddagger$.

The configurational entropy at the glass transition can be well approximated according to $s_c(T_g) \approx 0.5 \cdot s_m$~\cite{RWLbarrier, LW_soft}, where $s_m$ is the melting entropy of the corresponding crystal. The quantity $-(\partial \ln K/\partial \ln T)_{T_g}$ in the melt is generically around unity in chalcogenides~\cite{GADAUD2003146}; we set $(\partial \ln K/\partial \ln T)_{T_g}=-1$ for concreteness, see also SI.

\begin{table}
\caption{Jump in the apparent activation energy for electrical conductivity near the glass transition: Experiment from \cref{Eq:cond} (exp) and the present prediction (th).}
\begin{center}
\begin{tabular}{|c|c|c|c|c|}
\hline 
material & $\Delta E^\ddagger$(exp), eV & $\Delta E^\ddagger$(th), eV\\ 
\hline\hline
As$_{2}$Se$_{3}$ &$0.19 \pm 0.045$ & $0.081 \dots 0.22$ \\  
\hline
As$_{2}$S$_{3}$ &
$0.24 \pm 0.09$& $0.093$\\  \hline
\end{tabular}
\end{center}
\label{Tab:Eac}
\end{table}

The range of predicted values of $\Delta E^\ddagger$ for the {\em selenide}, see Table~\ref{Tab:Eac}, corresponds to the range of heights of the $\Delta c_p$ peaks reported in a separate, calorimetric study~\cite{doi:10.1111/j.1151-2916.1977.tb15488.x}. (The lowest peak indeed corresponds to the least-aged sample.) We see that the theoretically predicted low bound on $\Delta E^\ddagger$, as pertinent to non-aged samples, is indeed lower than the experimental values of $\Delta E^\ddagger$. On the other hand, the prediction that would pertain to the most-aged sample in Ref~\cite{doi:10.1111/j.1151-2916.1977.tb15488.x} is comparable to, if not greater than the experimental estimate from  Ref.~\cite{Davey1983}. This suggests the samples from Ref~\cite{Davey1983} had aged some. The  calorimetry data we found for the {\em sulfide}~\cite{WAGNER1998} do not exhibit a peak, thus indicating the calorimetry was done on a freshly made sample. At the same time, the predicted lower bound on $\Delta E^\ddagger$ for the sulfide is less than the experimental value to an extent that is similar to the selenide, which suggests the conductivity of the sulfide was likewise measured on an aged sample. 

When the peak is absent---corresponding to the equality in Eq.~(\ref{cpineq})---one may connect $\Delta E^\ddagger$ to the conventional fragility coefficient $m \equiv \partial \log_{10} \tau/\partial (T_g/T)|_{T_g}$, where $\tau_\alpha = \tau_{\alpha, 0} e^{F^\ddagger/T}$ is the $\alpha$-relaxation time and $F^\ddagger$ the corresponding free energy barrier. Since $F^\ddagger \approx K k_B/4 s_c$~\cite{LRactivated}, one obtains:
\begin{equation}\label{m}
\frac{m}{\log_{10} e} \approx \left[\frac{\Delta c_p(T_g)}{s_c(T_g)} + 1 - \left( \frac{\partial \ln K}{\partial \ln T} \right)_{T_g^+} \right] \frac{F^\ddagger(T_g)}{T_g}.
\end{equation}
Eq.~(\ref{cpineq}) then implies: 
\begin{equation} \label{DEm}
\frac{\Delta E^\ddagger}{k_B T_g/3} \gtrsim \frac{2 \, T_g}{F^\ddagger(T_g) \log_{10} e} \: m- 1 + \left( \frac{\partial \ln K}{\partial \ln T} \right)_{T_g^+}.
\end{equation}
The dynamical range of the glassy melt, $[F^\ddagger(T_g)/T_g]  \log_{10} e = \log_{10} [\tau(T_g)/\tau_0]$, is certainly less than $16$ but, most likely, no less than $10$. Thus the quantity multiplying $m$ in the equation above is greater than $1/8$ or so. For known substances $20 \lesssim m \lesssim 120$~\cite{L_AP}, which implies the jump $\Delta E^\ddagger$ should be distributed, among different substances, within the range $0.3 \cdot k_B T_g \dots 10 \cdot k_B T_g$ or so. Other relations connecting $\Delta E^\ddagger$ to material properties can be written, see SI. 

The optical gap itself exhibits a temperature dependence~\cite{LKgap}, which is approximately linear around $T_g$. This will contribute to the apparent pre-exponential factor in Eq.~(\ref{Eq:cond2}). We estimate the overall apparent prefactor, see SI, to be around $10^3 \ldots 10^4$~S/cm, consistent with Fig.~\ref{sigmaT}. The prefactor in Emin's scenario is somewhat greater but is comparable to the present prediction. Emin's scenario {\em also} predicts a non-vanishing $\Delta E^\ddagger$ since a melt expands more readily with temperature than the respective glass, but this volumetric effect is two orders of magnitude weaker than the prediction in Eq.~(\ref{Edjump}), see SI. 

A separate set of experimental challenges difficulties come about when the electrode material diffuses into the sample, thus artificially enhancing the conductivity, see SI for a detailed discussion. In any event, we hope this study will stimulate further experiments to test the present microscopic picture.

\section*{Final remarks}

We have argued that in glassy semiconductors with a spatially varying bond strength, the dominant charge carriers are polarons each bound to an intimate pair of topological defects. The defects are also intrinsically present as standalone entities; they are charged and spatially modulate the potential energy sensed by the charge carriers so as to lower the spatial density of states from the usually assumed atomic length to a larger, nanoscopic length $\xi$. The length $\xi$ reflects the cooperativity of activated transport in glassy melts. Thus conductivity measurements provide an unexpected venue to quantify the mechanism of the glass transition, a topic of much current interest. 

The present picture applies to glassy materials that host charge-density waves~\cite{ZL_JCP, LL1}. All known glassy semiconductors appear to fit into this category, whereby the bond order varies within unity while the magnitude of electronegativity variation is modest.  Disordered semiconductors that are made by deposition---or using other non-equilibrium methods---may or may not house a charge-density wave (CDW). For instance, amorphous silicon films exhibit (distorted) tetrahedral bonding  whose saturation is spatially uniform, and thus lack CDWs. At the same time silicon does not vitrify readily in the first place. Last but not least, we note that the present discussion of the dominant carrier applies to semiconductors that host a CDW, irrespective of whether the material is strictly periodic or not. Consequently, we expect topological polarons to dominate conductivity in intermetallic compounds whether a regular lattice and strict positional order are in place or not. This can be contrasted with Si, Ge, GaAs and other systems exhibiting a spatially uniform bond strength, where we expect the carriers to be extended wavepackets that conform to Migdal's theorem. On the other hand, the present argument on the {\em density of states} applies  only to glassy melts and frozen glasses. Thus the present results indicate that periodic materials that host CDWs would {\em also} conduct electricity via topological polarons, but the density of states is likely determined by other, system-specific factors.

{\bf Acknowledgments}: V.~L. thanks David Emin for insightful conversations. We gratefully acknowledge the support by the NSF Grants CHE-1465125 and CHE-1956389, the Welch Foundation Grant E-1765, and a grant from the Texas Center for Superconductivity at the University of Houston.
 
{\bf Significance Statement}: Constituent elements of a major class of semiconductors, exemplified by phase-change-memory (PCM) materials, exhibit ambiguous bonding preferences. This allows one to manipulate optoelectronic properties of these materials by inducing transitions between their ordered and disordered phases. Alongside, the coordination patterns around individual atoms switch in a discrete fashion. We show that in the presence of such bond switching, electricity must be carried by a special particle that is radically distinct from charge carriers found in classic semiconductor systems exemplified by crystalline silicon. In an independent test, the present theory predicts that the apparent activation energy for electrical conduction in glassy semiconductors should exhibit a discrete jump at the glass transition. The predicted values of the jump agree with existing measurements.}

%


\clearpage
\onecolumngrid
\begin{center}
\textbf{\large Supplementary Material}
\end{center}

\setcounter{equation}{0}
\setcounter{figure}{0}
\setcounter{table}{0}
\makeatletter
\renewcommand{\theequation}{S\arabic{equation}}
\renewcommand{\thefigure}{S\arabic{figure}}
\renewcommand{\bibnumfmt}[1]{[S#1]}
\renewcommand{\citenumfont}[1]{S#1}
%
\section{The topological midgap state and polaron solution of the Su-Schrieffer-Heeger Hamiltonian}
We describe the topological defects and the accompanying electronic states that arise in Peierls-dimerized chains using the Su-Schrieffer-Heeger Hamiltonian~\cite{RevModPhys.60.781-sub}:
\begin{equation}\label{Eq:SSH}
\Matr{H} = \sum\limits_{n, s}\Bigl[ - t^{}_{n+1, n}(\crn{c}_{n, s}\anh{c}_{n+1, s} + \crn{c}_{n+1, s}\anh{c}_{n, s}) + \epsilon^{}_n\crn{c}_{n, s}\anh{c}_{n, s}\Bigr] + \Matr{H}_{lat}.
\end{equation} 
Here $\crn{c}_{n, s}$ ($\anh{c}_{n, s}$) creates (annihilates) an electron with spin $s=\pm 1/2$ at site $n$.  The lattice component of the full energy describes, at a quadratic level, inter-site interactions as they would be in the absence of frontier electrons: 
\begin{equation}\label{Eq:Hlat}
\Matr{H}_{lat} = \frac{1}{2}\sum_{n}\left[m_a \dot{u}^2_n + \kappa(u^{}_{n+1} - u^{}_n)^2\right],
\end{equation}
where $\kappa$ is the spring constant, $m_a$ the atomic mass, and $u_n$ the displacements of the site off the locations they would have in a chain with undeformed springs. 

We follow SSH by adopting the following parametrization of the hopping matrix element $t_{n+1, n}$ as a function of the inter-site separation:  
\begin{equation}\label{Eq:Hopping}
t_{n+1, n} = t_0 - \alpha |u_{n+1} - u_{n}|,
\end{equation}
which is a linear approximation for the actual, approximately exponential dependence of inter-site matrix elements.  

The on-site energies $\epsilon_n$ are introduced, as in Refs.~\cite{PhysRevLett.49.1455-sup, ZL_JCP-sup}, to model a spatial distribution of the electronegativity. For concreteness we adopt a perfectly alternating pattern of electronegativity variation along the chain: 
\begin{equation}
    \epsilon_n = (-1)^{n}\epsilon.
\end{equation}
This is the case considered and solved by Rice and Mele~\cite{PhysRevLett.49.1455-sup}, who showed that the system will become Peierls-{\em stable} when the electronegativity variation exceeds a certain threshold value. In the latter case, the chain does not spontaneously form a charge-density wave (CDW) but, instead, represents an ionic insulator. Thus we adopt here a sufficiently small value of $\epsilon$ such that the (doubly-degenerate) ground state of our chain, at half-filling, is a perfectly Peierls-dimerized chain. Denote the ground state value of the staggered displacement 
\begin{equation}
   \phi_n \equiv  (-1)^{n}u_{n},
\end{equation}
with $u_0$. (This quantity is doubly degenerate.) Denote the values of the larger and smaller overlap integrals with $t_+$ and $t_-$, respectively. This system does host a charge-density wave, whose magnitude scales approximately linearly with the hopping-element differential $|t_+ - t_-|$. 
 
We next choose concrete values for the parameters $t_0$, $\alpha$, and $\kappa$ that are roughly consistent with mechanical and electronic properties of the chalcogenides of interest, such as arsenic selenide or sulphide.  The optical gap is about 1.8~$\mathrm{eV}$ for As$_{2}$Se$_{3}$, $2.2$~$\mathrm{eV}$ for As$_{2}$S$_{3}$. For concreteness, we set the insulating gap in the ground state of our dimerized chain at 2~$\mathrm{eV}$, at $\epsilon=0$. This provides a constraint for the hopping matrix element differential: $t_{+} - t_{-} = 4\alpha u_{0} =  E_{g}/2$. An additional constraint on the parameters can be obtained by noting that the zone width in the SSH Hamiltonian, at $\epsilon=0$, is approximately equal to \cite{RevModPhys.60.781-sub}: 
\begin{equation}
    \kappa/\alpha^{2} \approx -\frac{4\pi}{3}\left[1 + \ln\left(\frac{E_{g}}{16 t_{0}}\right)\right],
\end{equation}
while the width of the valence zone in As$_2$Se$_3$ is calculated to be around 6~$\mathrm{eV}$ \cite{LL2-sup}. This, then, imposes bounds on the choice of the parameter $\alpha$ since the spring constant $\kappa$ should be in the range $1 \ldots 10$~$\mathrm{eV}$$/$\AA~\cite{LW-sup}. In what follows, we adopt values $t_{-} = 2.5$~$\mathrm{eV}$ and $t_{+} = 3.5$~$\mathrm{eV}$.

Excitations in the SSH Hamiltonian can be quite complicated because they mutually couple the electronic and vibrational motions, even though there is no interaction {\em within} the respective individual sets of motions. To simplify reasoning, it is often useful to think of a dimerized chain in an ultra-local limit of a set of weakly interacting dimers~\cite{LL2-sup}, whereby $t_{-} \ll t_{+}$. As a concrete chemical implementation of this limit, imagine an even-numbered chain of hydrogen atoms. It is obvious that for its ground state, such a chain will break up into a set of H$_2$ dimers, so that the bonding within each dimer is covalent, while nearest-neighbor dimers are coupled but weakly, through a closed-shell interaction. In this extreme limit and $\epsilon=0$, it is obvious that a chain containing an odd number $n$ of atoms, in its ground state, will have precisely one excess weak or strong bond. Molecular realizations of the latter arrangement, as pertinent to chalcogenide alloys, can be found in Refs.~\cite{ZLMicro2-sup, GHL-sup, LL2-sup}. Such coordination defects can only be removed by changing the number of sites and, therefore, are topologically stable against vibrations. At the defect, the staggered displacement $\phi_n$ will exhibit a discrete sign reversal. The cases of an excess weak and strong bond can be thought of as an extra single site or an extra trimer, respectively~\cite{GHL-sup}. In either case, there will appear an edge-like, non-bonding orbital located inside the forbidden gap~\cite{LL2-sup}.

\begin{figure}[t]
\center
{\includegraphics[width= .5 \columnwidth]{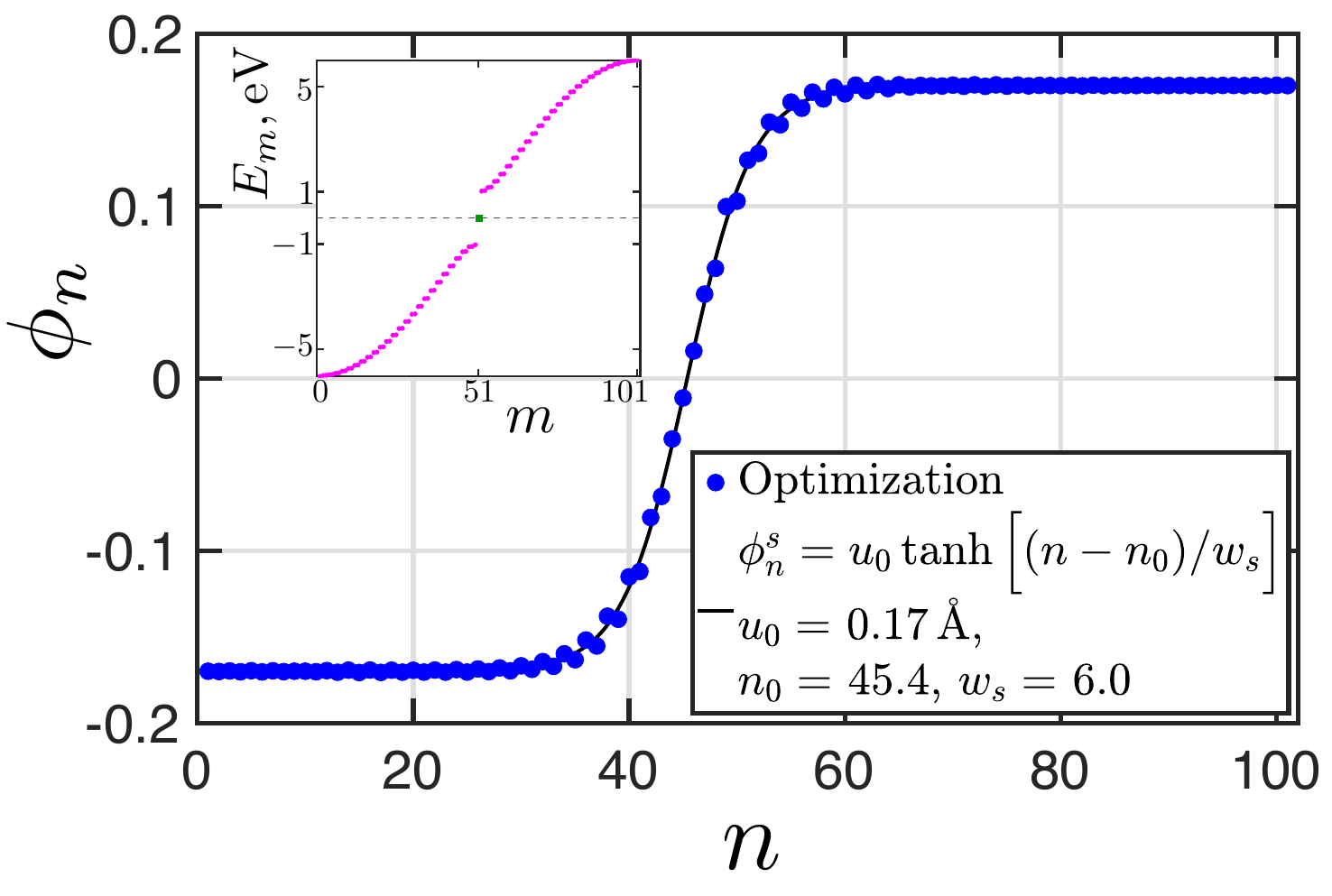}}
\caption{\label{Fig:Sol} Ground-state soliton solution of the SSH Hamiltonian: Staggered displacement $\phi_{n}$, as a function of site number $n$, is shown as a function of site $n$. The solid line is a fit using the form \cref{Eq:Soliton}. Inset: the full electronic spectrum. The midgap state, shown with the light-blue symbol, is strictly in the middle of the gap. $\kappa = 2.1$~$\mathrm{eV}$/\AA$^2$, $t_{0} = 3$~$\mathrm{eV}$, $\alpha = 1.5$~$\mathrm{eV}$/\AA. }
\end{figure}

When the disparity between the hopping elements $t_{-}$ and $t_{+}$ is not very large, the defect and the accompanying midgap state will persist, but the aforementioned discrete sign change will be replaced by a milder, sigmodal curve. In other words, the deviation from the perfect coordination pattern, as well as the accompanying lattice strain, are now distributed over a relatively extended region. At the center of the strained region, the bonds  will be intermediate, length- and strength-wise, between the weak and strong bond, respectively, of a dimerized chain~\cite{ZLMicro2-sup}. The width of the kink scales roughly inversely proportionally with the band gap~\cite{RevModPhys.60.781-sub, ZL_JCP-sup}. It is common to fit the sigmodal dependence using the following functional form: 
\begin{equation}\label{Eq:Soliton}
\phi^s_n = u_0\tanh\left(\frac{n - n_0}{w_s}\right).
\end{equation}
This form becomes exact in the continuum limit~\cite{PhysRevB.21.2388-sup, RevModPhys.60.781-sub}. The latter limit also drives home the solitonic nature of the coordination defect. The parameter $w_s$ can be thought of as the half-width of the soliton. 

For a chain at half-filling, the midgap state will be singly occupied. The resulting energy cost is about $0.62~E_g/2$ for a long chain; it tends asymptotically to $E_g/\pi$ in the continuous limit~\cite{PhysRevB.21.2388-sup}. We show a soliton in its vibrational ground state, at $\epsilon=0$, for a closed chain of 101 sites in \cref{Fig:Sol}. 

The topological solitons represent a family of gapped excitations of the chain. While we generated the soliton solution above by employing an odd-numbered chain---which drove home the soliton's topological nature---such solitons can be generated in extended chains irrespective of the parity of the total chain length. Indeed, following activation, a pair of defects can spontaneously emerge in a perfectly dimerized chain, one defect corresponding to an undercoordinated center, the other to an overcoordinated center~\cite{RevModPhys.60.781-sub, ZL_JCP-sup}. Imagine that the so emerged two defects fail to mutually annihilate but, instead, separate in space. An individual soliton or anti-soliton should each be considered as a standalone excitation, the same way we consider electrons and holes as standalone excitations. A standalone soliton will equilibrate vibrationally yet no restoring force from the chain will appear that attempts to remove the soliton; the latter can be removed only via annihilation with a separate solition of opposite polarity, as already alluded to. In addition, the soliton can travel unimpeded along the chain~\cite{RevModPhys.60.781-sub}.

\begin{figure}[t]
\center
\subfigure[]{\includegraphics[width= .5 \columnwidth]{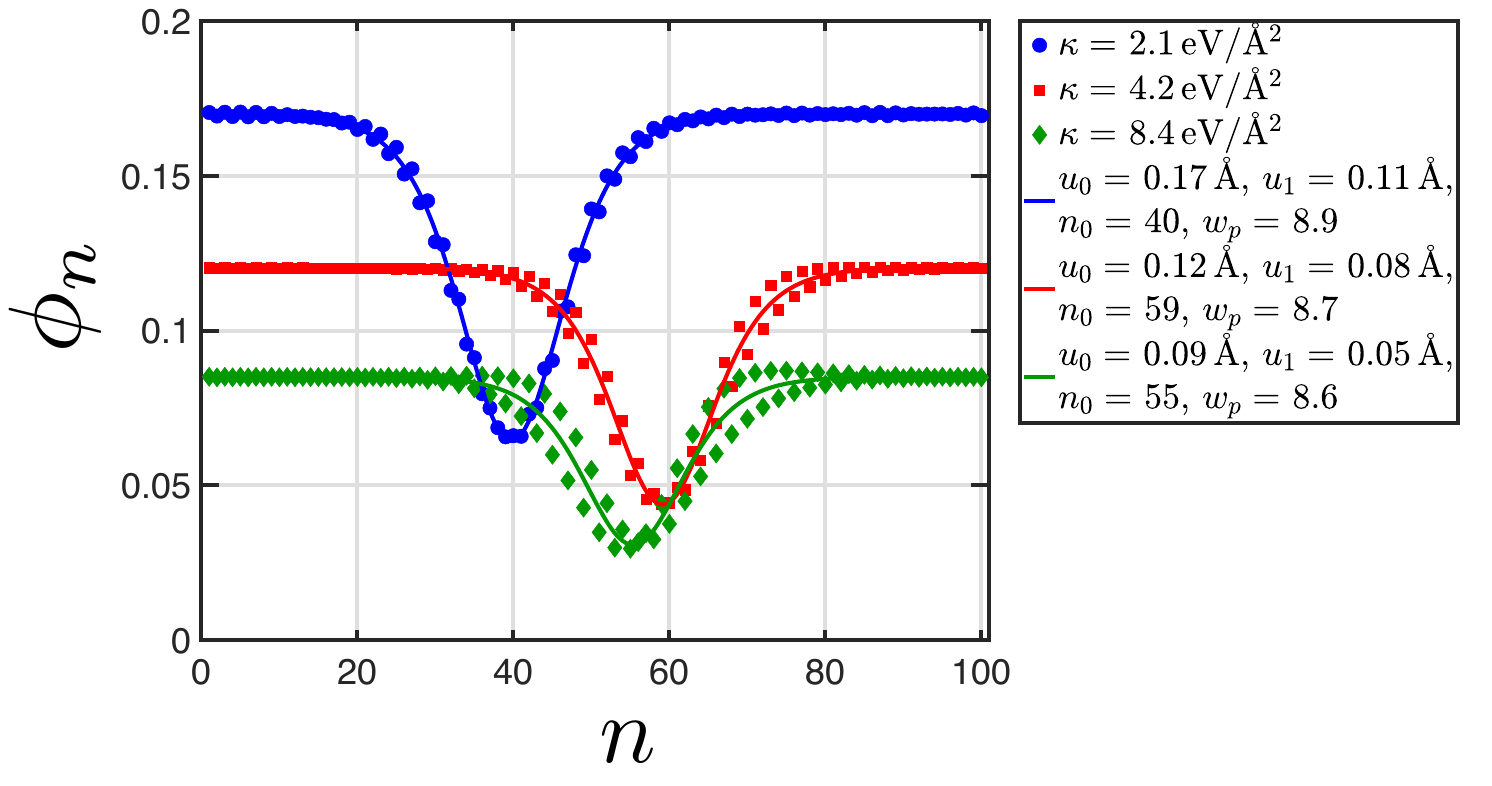}}\hfill
\subfigure[]{\includegraphics[width= .5 \columnwidth]{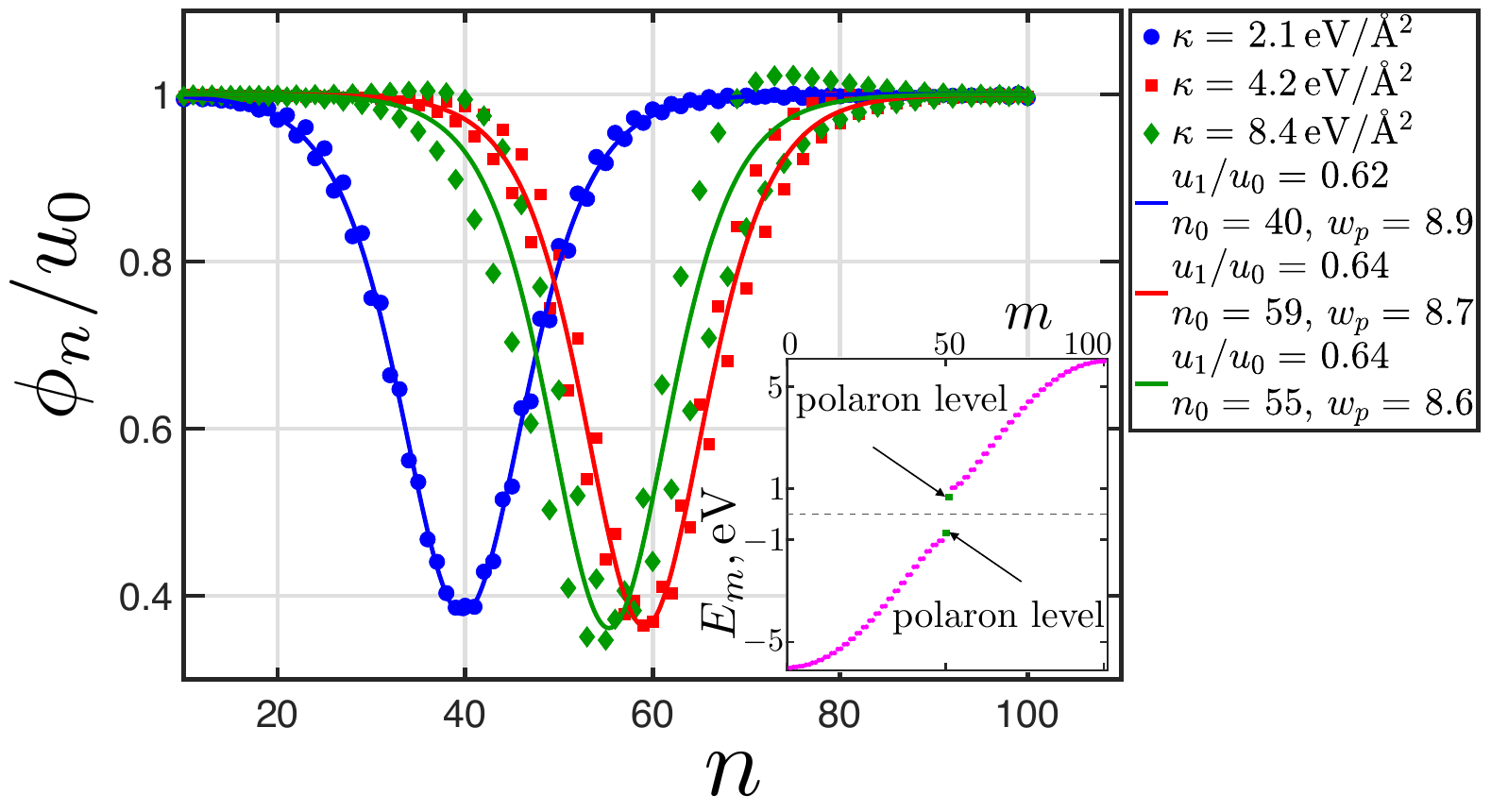}}
  \caption{\label{Fig:Pol} Ground state polaron solution of the SSH Hamiltonian (a) Staggered displacements for select values of $\kappa$. $\alpha = 1.5$ (blue circles), $2.1$ (red squares), $3.0$~$\mathrm{eV}$/\AA (green diamonds). The parameter $t_{0}=3$~$\mathrm{eV}$, as in \cref{Fig:Sol}. Solid lines show the fit using the functional form \cref{Eq:PolaronProfile}. (b) Main graph: Same as (a) but normalized to the maximum value. Inset: The electronic spectrum. The polaron levels are $\Delta E_{p}\approx 0.3$~$\mathrm{eV}$ from the mobility zones.  The gap is $E_{g} \approx 2.0$~$\mathrm{eV}$.}
\end{figure}

An alternative set of intrinsically gapped excitations for a vibrationally equilibrated chain come about when an electron/hole is added to the conduction/valence band. This type of excited state represents a polaron. It is common for electronically excited states to interact with the lattice and, in particular, with optical phonons~\cite{LKgap-sup}. Such interactions serve to stabilize charge-separated states, resulting in a Stokes shift in photoemission, among other things. Similarly, the total energy of a polaron in a Peierls-distorted chain will be lowered somewhat relative to the appropriate edge of the insulating gap, following vibrational relaxation of the chain~\cite{Su5626-sup, ISI:A1981MD41000002-sup, RevModPhys.60.781-sub}. In fact, the corresponding vibration should be classified largely as an optical vibrational mode: In the presence of a polaron, the stronger bond housing the polaron will elongate, as mentioned in the main text. This is an optical mode because it involves displacement within a unit cell of the dimerized chain. Alternatively, this displacement can be thought of an intimate soliton---anti-soliton pair because it attempts to create, next to each other, two nearest-neighbor short bonds, on the one hand, and two nearest-neighbor long bonds, on the other hand. This suggests that one may fit the resulting spatial profile of the staggered displacement using the spatial derivative of the solitonic profile from \cref{Eq:Soliton}, viz.:
\begin{equation}\label{Eq:PolaronProfile}
\phi^p_n = u_{0} - u_{1}\sech^{2}\left[\frac{n - n_{0}}{w_p}\right].
\end{equation}   

We note that at the SSH level, the energy of the polaron level, relative to the middle of the gap, depends on the parameters through the gap width $E_g$ alone, at $\epsilon=0$~\cite{Emin1980-sup, ISI:A1981MD41000002-sup, CAMPBELL1982297-sup}:
\begin{equation}\label{Eq:PolaronLevel}
E_p^{\pm} = \pm\frac{\sqrt{2}}{2}\frac{E_g}{2},\quad \Delta E_{p} = \frac{E_{g}}{2}\left(1 - \frac{\sqrt{2}}{2}\right),
\end{equation}  
where $\Delta E_{p}$ is  the splitting of polaron levels off the edges of the electronic bands. Here we illustrate the polaron solution, at $\epsilon=0$. We have optimized a chain of length $N = 100$ filled with $101$ electrons. The result is shown in \cref{Fig:Pol}. We find that $\Delta E_{p}\approx 0.3$~$\mathrm{eV}$ for the gap $E_{g}\approx 2.0$~$\mathrm{eV}$, in agreement with \cref{Eq:PolaronLevel}. We have observed that the polaron width is independent of the spring constant as long as the electron-phonon coupling $\alpha$ is adjusted so as to maintain the gap fixed.

In the main text, we illustrate the soliton and polaron solutions at a non-vanishing value of $\epsilon = 0.1$~$\mathrm{eV}$, as would  be appropriate for chalcogenides. The rest of the parameters are as follows: $t_0 = 3.0$~$\mathrm{eV}$, $\kappa = 2.1$~$\mathrm{eV}$/\AA, and $\alpha = 1.5$~$\mathrm{eV}$/\AA. Both solutions are qualitatively and quantitatively similar to solutions obtained at $\epsilon=0$, consistent with the notion that these excitations come about owing to the charge-density wave, while the non-vanishing electronegativity variation is a perturbation.

\section{\texorpdfstring{$\Delta E^\ddagger$}{Lg} Estimates}

\subsection{Experimental data and fits}

\begin{figure}[ht]
\center
\begin{tabular}{cc}
 \includegraphics[scale = .51]{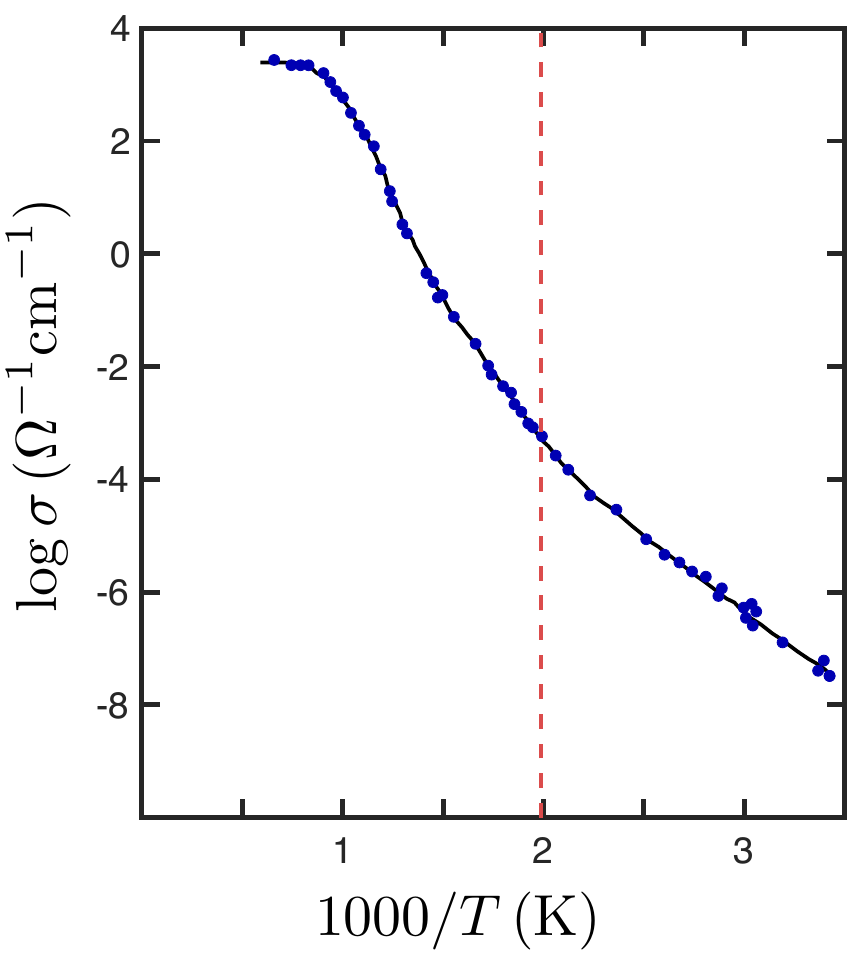}    &  \includegraphics[scale = .51]{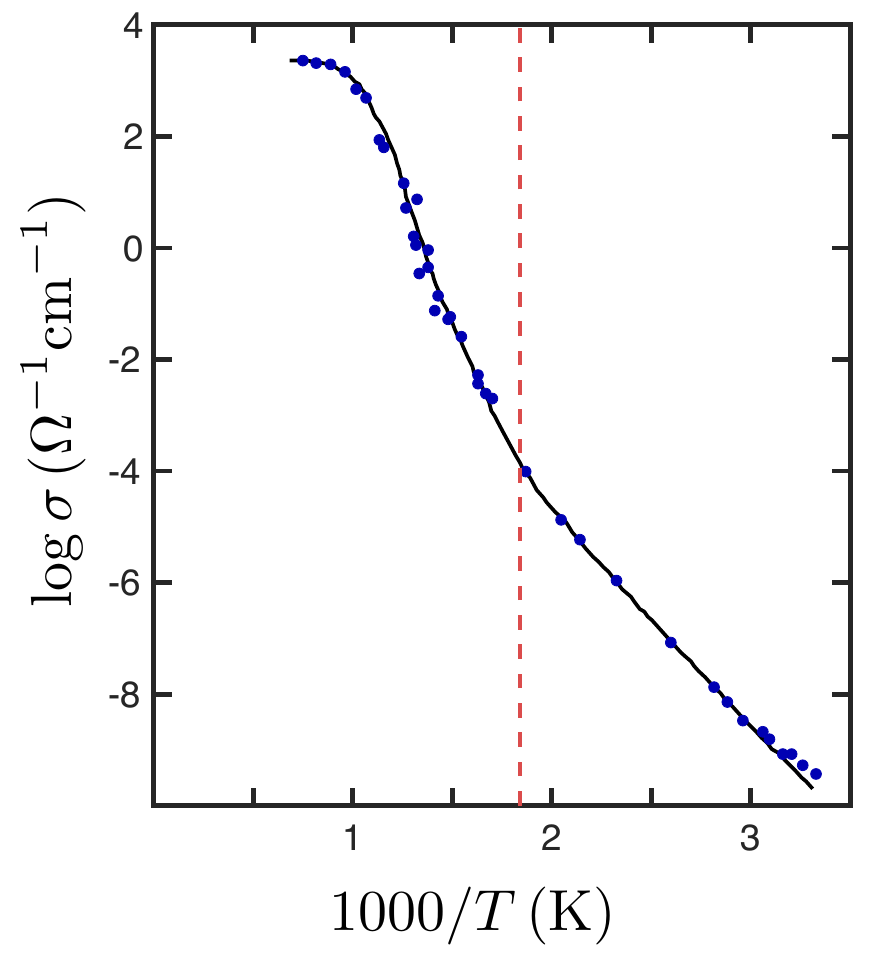}
\end{tabular}
\caption{\label{Fig:sigmaQualitative} Electric conductivity of As$_{30}$Te$_{48}$Ge$_{10}$Si$_{12}$ (left) and S$_{15}$Ge$_{33}$As$_{27}$Te$_{25}$ (right), according to Davey and Baker~\cite{Davey1983-sup}. The dashed vertical lines indicate the respective glass transition temperatures. The lines are provided as guides to the eye, similarly to Ref.~\cite{Davey1983-sup}.}
\end{figure}

We begin by highlighting essential qualitative features of the temperature dependence of the electrical conductivity in phase-change materials as exemplified by two quaternary compounds in Fig.~\ref{Fig:sigmaQualitative}.
The graph on the left displays three well-sampled regimes; see also an early discussion by Edmond~\cite{Edmond-sup}. The Arrhenius-like regime below the glass transition temperature $1000/T \gtrsim 2.0$---call it (I)---has an activation energy that is numerically close to a half of the optical gap. Immediately above the glass transition $1.3 \lesssim 1000/T \lesssim 2.0$---call it regime (II)---an Arrhenius-like dependence is {\em also} observed, over three orders of magnitude or so, but with a somewhat larger figure for the apparent activation energy. At the higher yet temperatures, $1000/T \lesssim 1.3$, one observes a greater yet rate of increase of the conductivity followed by an eventual leveling off at values typical for a poor metal. This latter regime---call it (III)---pertains to the metal-insulator transition~\cite{LKgap-sup} and is outside the scope of the present work. The r.h.s. panel of Fig.~\ref{Fig:sigmaQualitative} is presented here to illustrate yet another distinct physical regime that comes about in deeply frozen glasses, manifesting itself in the r.h.s. panel as a deviation from an Arrhenius-like dependence at $1000/T \gtrsim 3.0$. This lowest-$T$ regime, often ascribed to variable range hopping~\cite{Mott1993-sup}, is {\em also} outside the scope of the present work and will not be mentioned in what follows.

\begin{figure}[t]
\center
\begin{tabular}{ccc}
 \includegraphics[scale = .38]{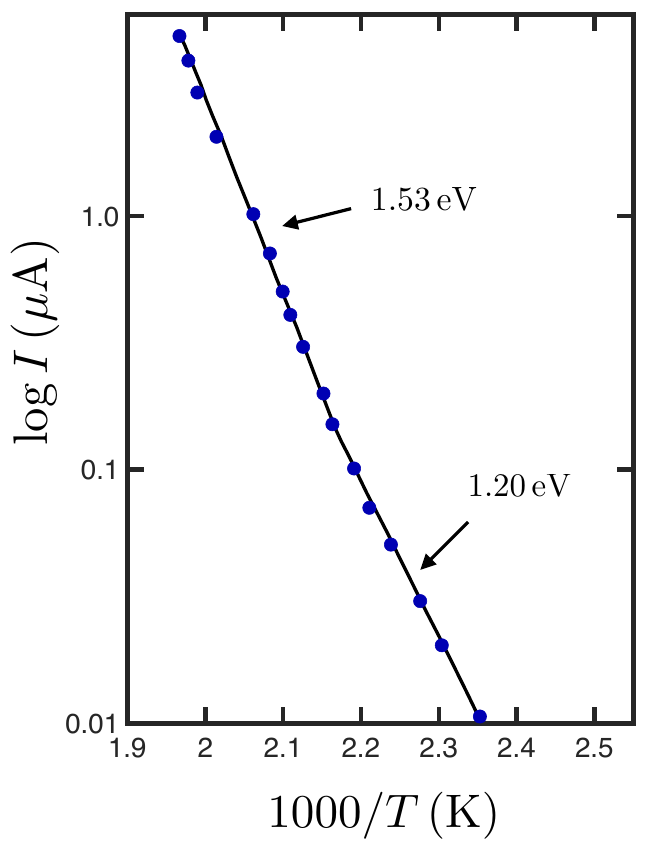}    & \includegraphics[scale = .38]{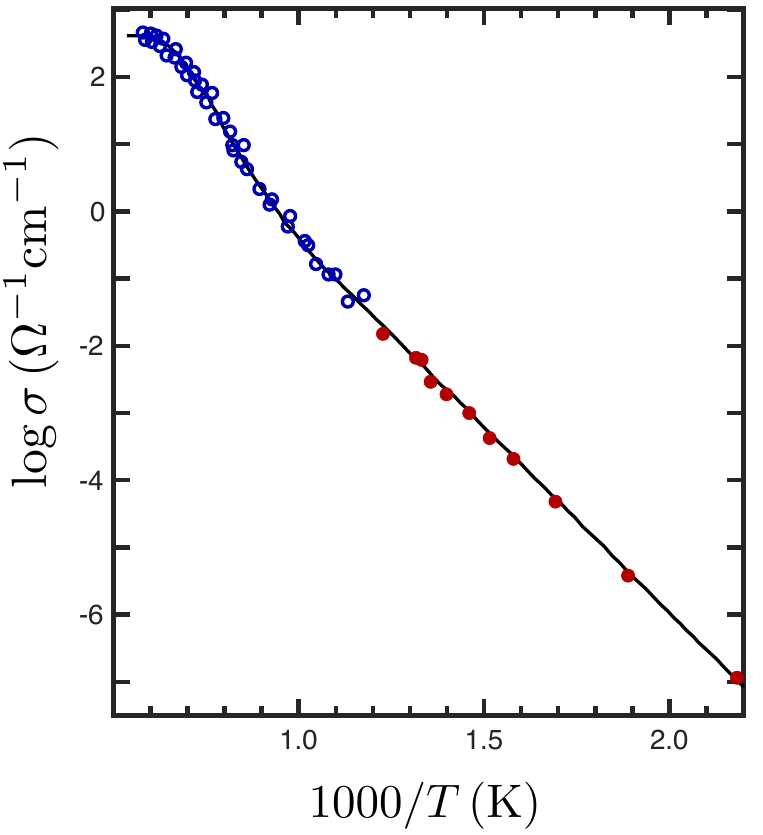}    & \includegraphics[scale = .38]{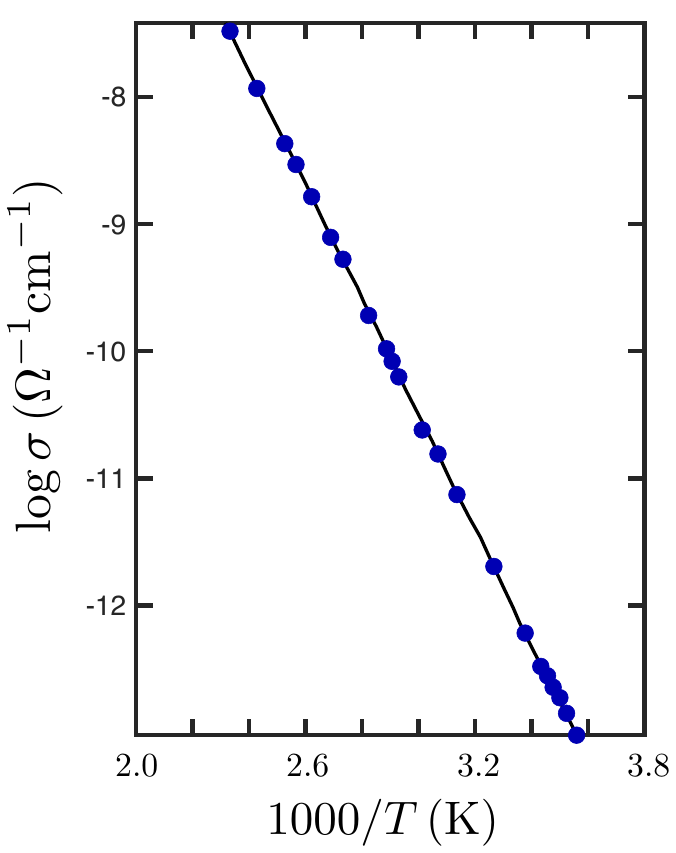}
\end{tabular}
\caption{\label{Fig:condFits} \textit{Left}: $I(T)$ of As$_2$S$_3$, according to Bobb et al. \cite{Bobb1975techreport-sup}.  \textit{Center}: $\sigma(T)$ for the glassy As$_2$Se$_3$ {\em melt}, according to Webb and Baker~\cite{WebbBaker1972-sup}. The red dots indicate which points were used for the present fits.  \textit{Right}: $\sigma(T)$ for the frozen As$_2$Se$_3$ {\em glass}, according to Seager and Quinn~\cite{SeagerQuinn-sup}. The lines are provided as guides to the eye, similarly to the source articles.}
\end{figure}

The present study focuses exclusively on regimes (I) and (II). We have found only two substances for which both conductivity and calorimetry data were available, viz., the stoiochiometric binary compounds As$_2$S$_3$ and As$_2$Se$_3$. In Fig.~\ref{Fig:condFits}, we show the actual conductivity data used for the present fits. In the case of the selenide melt, we added red symbols on top of the facsimile to indicate the points that we chose to represent regime (II).   
We fit the  $\ln \sigma(T)$ vs. $1/T$ curves for each substance using a set of two straight lines, one line pertaining to $T < T_g$ (regime (I)), the other to $T > T_g$ (regime (II)). The corresponding negative slopes are denoted with $E^{<}$ and $E^{>}$, respectively. (The error of the slopes is evaluated within the $67\%$-confidence interval.) The slope discontinuity is consequently evaluated according to: 
\begin{equation}
   \Delta E^\ddagger_{\rm exp} \equiv E^> - E^<. 
\end{equation}
\begin{center}
\begin{tabular}{|c|c|c|c|}
\hline
Material &  $E^{<}$, $\mathrm{eV}$&$E^{>}$, $\mathrm{eV}$ &$\Delta E^\ddagger_{\rm exp}$, $\mathrm{eV}$\\
\hline
As$_{2}$Se$_{3}$ \cite{Davey1983-sup,WebbBaker1972-sup} &$0.90\pm0.002$ &$1.09\pm0.045$ &$0.19\pm0.045$ \\
\hline
As$_{2}$S$_{3}$ \cite{Bobb1975techreport-sup} &$1.26\pm0.07$ &$1.50\pm0.06$ &$0.24\pm0.09$ \\
\hline
\end{tabular}
\end{center}

\begin{figure}[t]
\center
\includegraphics[scale = .5]{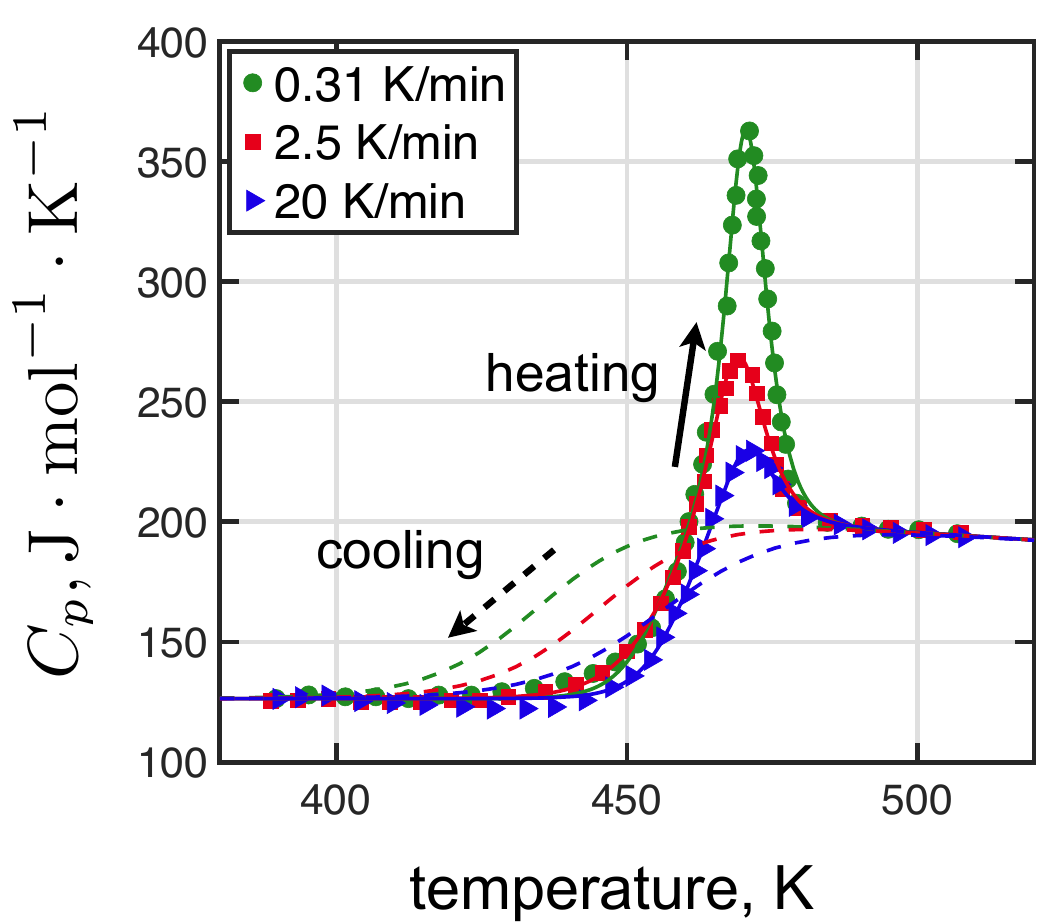}
\caption{\label{Fig:Cp_HeatCoolAll} Symbols: The excess liquid heat capacity relative to the crystal $C_{p}(T)$ vs $T$ for As$_{2}$Se$_{3}$, for three distinct cooling protocols differing by the quench rate of the melt, as indicated in the legend. The heating is all at the same rate of 10~$\mathrm{K/min}$. The smooth lines are fits from Eqs.~\cref{Eq:CpHeat,Eq:ProbHeat} and \cref{Eq:CpCool,Eq:ProbCool}. Original data from Ref.~\cite{doi:10.1111/j.1151-2916.1977.tb15488.x-sup}.}
\end{figure}

There appears to be a lack of agreement in the literature regarding the possibility of the electrode material diffusing inside the sample, which might substantially skew the conductivity data toward higher values. Bobb et al.~\cite{Bobb1975techreport-sup} state that ``silver readily diffuses inside the sample lowering the resistivity.'' Seager and Quinn~\cite{SeagerQuinn-sup} state to the contrary while, at the same time, reporting substantially higher conductivities in the As$_2$S$_3$ glass, as does Borisova~\cite{Borisova1981-sup}. We have chosen to use Bobb et al.~\cite{Bobb1975techreport-sup} data for the sulfide. For the selenide glass, direct data on the conductivity in the frozen glass are available only from Seager and Quinn~\cite{SeagerQuinn-sup}. These data are consistent with the figures of conductivity provided by Davey and Baker~\cite{Davey1983-sup}, which they infer from Edmond's optical absorption data~\cite{Edmond-sup}. Although indirect, the optical gap method has the advantage of not being subject to uncertainties stemming from sample-electrode interaction. In any event, the Seager and Quinn~\cite{SeagerQuinn-sup} sulfide data, together with the present theory, seem to suggest a value for $T_g$ that is somewhat higher than its calorimetric value $460$~K or so. Calorimetric $T_g$'s are often tied to the inflection point on the temperature dependence of the excess liquid heat capacity relative to the corresponding crystal, see for instance Fig.~\ref{Fig:Cp_HeatCoolAll}.

Conductivity data consistent with a jump in the apparent activation energy are available for many systems, see for instance Ref.~\cite{Borisova1981-sup}. Many of these systems, including phosphorus chalcogenides and various non-stoichiometric selenides appear to be chemicallly unstable, however, nor could we find descriptions of sample preparation. None of the $\sigma(T)$ references we have found contained information on the quenching rate used to make the respective glass either. The present estimates suggest the samples had been aged, as we point out in the main text. This is consistent with Bobb et al.s~\cite{Bobb1975techreport-sup} statement that the sample had been acquired from elsewhere. To summarize, the present discussion that systems most suitable for testing the present picture should (a) be stable against ordering or separation; (b) exhibit little interaction with the electrode material; (c) desirably have a gap not too large so as to make it easier to measure the conductivity.

We proceed next to the calorimetry. A separate, exclusively calorimetric study due to Eastel et al.~\cite{doi:10.1111/j.1151-2916.1977.tb15488.x-sup} reports a range of preparation protocols for As$_2$Se$_3$. We replot their heat capacity data per stoichiometric unit in Fig.~\ref{Fig:Cp_HeatCoolAll}. For typographical convenience, we drop the ``$\Delta$'' in the notation for the excess heat capacity $\Delta C_p$ of the liquid relative to the crystal. In the same figure, we also provide smooth fits that employ ad hoc functional forms.  For heating (solid lines), we have:
\begin{equation}\label{Eq:CpHeat}
C_{p}^{\rm heat}(T) = C_{p}^{\rm g} + \left(C_{p}^{\rm gr} - C_{p}^{\rm g}\right)\mathcal{F}_{\rm trial}^{\rm heat}(T),
\end{equation} 
where the trial function is given by  
\begin{equation}\label{Eq:ProbHeat}
\mathcal{F}_{\rm trial}^{\rm heat}(x) = \frac{1}{2} + \frac{1}{2}\tanh\left(a[x - x_{1}]\right) + \eta \, \sech\left(b[x - x_{2}]\right).
\end{equation}
We fix the specific heat at $T=395\,\mathrm{K}$ at the value $C_{p}^{\rm g} = 127\,\mathrm{J/(mol\cdot \mathrm{K})}$ and at $T=505\,\mathrm{K}$ at the value $C_{p}^{\rm gr} = 203\,\mathrm{J/(mol\cdot K)}$. Subsequently, $a$, $b$, $\eta$, $x_{1}$, $x_{2}$ are determined by fitting. For cooling we use the form:
\begin{equation}\label{Eq:CpCool}
C_{p}^{\rm cool}(T) = C_{p}^{\rm g} + \left(C_{p}^{\rm gr} - C_{p}^{\rm g}\right)\mathcal{F}_{\rm trial}^{\rm cool}(T),
\end{equation} 
where trial function is given by
\begin{equation}\label{Eq:ProbCool}
\mathcal{F}_{\rm trial}^{\rm cool}(x) = \frac{1}{a\exp\left\{-b(x - x_{0})\right\} + 1},
\end{equation}
with $a$, $b$, $x_{0}$ as fitting parameters.

We list below the resulting values of the peak values of the excess heat capacity of the melt relative to the glass: 
\begin{center}
\begin{tabular}{|c|c|}
\hline
Cooling rate, $\mathrm{K/min}$ & $\Delta c_{p}^{\rm max}$, $\mathrm{J/(mol\cdot K)}$ \\
\hline\hline
$20$ & $95$ \\
\hline
$2.5$ & $133$ \\
\hline
$0.31$ & $228$ \\
\hline
\end{tabular}
\end{center}
We use the cooling part of the protocol to infer $\Delta c_p(T_g)=65$~$\mathrm{J/(mol\cdot K)}$ for As$_2$Se$_3$, which is in agreement with Wagner et al.~\cite{WAGNER1998-sup}. The latter work also reports calorimetry data for the sulfide As$_2$S$_3$, viz.,  $\Delta c_p(T_g)=62$~$\mathrm{J/(mol\cdot K)}$ and no discernible peak. The aforementioned calorimetry data are compiled in the Table below and serve as input in evaluating the low bound in Eq.~4 of main text. 

\begin{center}
\begin{tabular}{|c|c|c|c|c|}
\hline
material & $\Delta c_p(T_g)$, $\mathrm{J/(mol\cdot K)}$  & $s_m/2$, $\mathrm{J/K}$ &  $T_{\rm g}$,~$\mathrm{K}$ & $T_{\rm K}$,~$\mathrm{K}$ \\
\hline
As$_{2}$Se$_{3}$ & $65$ & $31.4$ & $460$  & $236\pm 10$ \\
\hline
As$_{2}$S$_{3}$ & $62$ & $24.5$ & $465$  & $265$  \\
\hline
\end{tabular}
\end{center}

We approximate the value $s_c(T_g)$ of the configurational entropy as $s_{c}(T_{g}) \approx 0.5 \cdot s_m$, where $s_m$ is the entropy of melting of the corresponding crystal. This approximation is based on two established notions: On the one hand, the configurational entropy, at the glass transition, varies within a modest range~\cite{XW-sup, LW_soft-sup, RWLbarrier-sup, LRactivated-sup} $0.8 \ldots 0.9$~$k_B$ per bead. On the other hand, the number of beads for a compound can be determined by calibrating its fusion entropy against the fusion entropy per particle of the Lennard-Jones liquid~\cite{LW_soft-sup, RWLbarrier-sup}, viz. $s_m^\text{(LJ)} = 1.67$~$k_B$. Thus we use $s_c(T_g) \approx 0.5 s_m = 0.5 \Delta h_m/T_m$, where $\Delta h_m$ is the fusion enthalpy ($40.8$~$\mathrm{kJ/mol}$ (As$_2$Se$_3$), $28.7$~$\mathrm{kJ/mol}$ (As$_2$S$_3$) \cite{CRC-sup}) and $T_m$ is the melting temperature ($650\,\mathrm{K}$ (As$_2$Se$_3$), $585\,\mathrm{K}$ (As$_2$S$_3$) \cite{CRC-sup}). We list the resulting values of $s_m/2$ in the Table above, as well as values of $T_g$~\cite{SeagerQuinn-sup, https://doi.org/10.1002/pssb.19640070302-sup} and the Kauzmann temperature $T_K$, as reported in Ref.~\cite{PhysRevLett.64.1549-sup}, to be used in evaluating \cref{Edjump3Supple,Edjump4Supple} below.

\subsection{\texorpdfstring{$\Delta E^\ddagger$}{Lg} predictions: Numerical estimates}

We do not have data on the dimensionless rate of decrease of the bulk modulus with temperature  $(\partial \ln K/\partial \ln T)_{T_g^+}$ for either As$_2$Se$_3$ or As$_2$S$_3$. The quantity $(\partial \ln K/\partial \ln T)_{T_g^+}$ is, however, not expected to be too different from negative unity, judging from data on a different intermetallic compound~\cite{GADAUD2003146-sup}. Thus we estimate Eq.~(4) for two specific values of $(\partial \ln K/\partial \ln T)_{T_g^+}$ viz. $0$ and $-1$, respectively,  $\Delta E^\ddagger$, to get a quantitative feel for the significance of the thermally-induced changes of the elastic response. Thus we compute two quantities, as pertinent to Eq.~(4) of the main text: 
\begin{equation}\label{Edjump1Supple}
\Delta E^\ddagger_{1}  = \frac{k_B T_g}{3} \left[\frac{2 \Delta c_p^\text{(max)}}{s_c(T_g)} + 1\right].
\end{equation}
\begin{equation}\label{Edjump2Supple}
\Delta E^\ddagger_{2}  = \frac{k_B T_g}{3} \!\! \left[\frac{2 \Delta c_p^\text{(max)}}{s_c(T_g)} + 1 - \left( \frac{\partial \ln K}{\partial \ln T} \right)_{T_g^+} \right].
\end{equation}

As an alternative to the fragility-based estimate in Eq.~(7) of the main text, one may also use the conventional parameterization~\cite{RichertAngell-sup, LW_soft-sup} $s_{c}(T) \propto  (1/T_K - 1/T) $, where $T_K$ is the so called Kauzmann temperature. For known glassformers, $0.5 \lesssim  T_K/T_g \lesssim 0.9$~\cite{WangAngell, LW_soft-sup}. Assuming $s_c = \text{const}$ at $T< T_g$, one obtains straightforwardly
\begin{equation}\label{EdjumpGEQsupple}
\frac{\Delta E^\ddagger}{k_B T_g/3} \geqslant
\frac{1 + T_K/T_g}{1 - T_K/T_g}.
\end{equation}
Thus one may evaluate the following two lower bounds on $\Delta E^\ddagger$:
\begin{equation}\label{Edjump3Supple}
\Delta E^\ddagger_{3} =
\frac{k_B T_g}{3}  \frac{1 + T_K/T_g}{1 - T_K/T_g}.
\end{equation}
and
\begin{equation}\label{Edjump4Supple}
\Delta E^\ddagger_{4} =
\frac{k_B T_g}{3}  \left[\frac{1 + T_K/T_g}{1 - T_K/T_g} - \left( \frac{\partial \ln K}{\partial \ln T} \right)_{T_g^+}\right] .
\end{equation}

Numerical estimates of the expressions in \cref{Edjump1Supple,Edjump2Supple,Edjump3Supple,Edjump4Supple} are given in the Table below:
\begin{center}
\begin{tabular}{|c|c|c|c|c|}
\hline
material & $\Delta E^{\ddagger}_{1}$, $\mathrm{eV}$&  $\Delta E^{\ddagger}_{2}$, $\mathrm{eV}$&$\Delta E^{\ddagger}_{3}$, $\mathrm{eV}$&$\Delta E^{\ddagger}_{4}$, $\mathrm{eV}$\\
\hline
As$_{2}$Se$_{3}$ & $0.068 \dots 0.20$ &$0.081\dots0.22$ &$0.041$ & $0.054$\\
\hline
As$_{2}$S$_{3}$ &  $0.080$  & $0.093$  & $0.049$ & $0.062$ \\
\hline
\end{tabular}
\end{center}

Of the four options, we regard the quantity $E^\ddagger_2$ from \cref{Edjump2Supple} as the most reliable estimate and, hence, present it as our predicted value in the main text. We see that not knowing the value of $(\partial \ln K/\ln T)$ translates to at most a 15\% uncertainty in the first place.  

Eqs.~(7) in the main text and \cref{Edjump3Supple,Edjump4Supple} are very simple algebraically but, at the same time, are subject to greater uncertainty than the $E^\ddagger_2$.  

Eq.~(7) only gives a lower bound and requires one to independently determine the actual, not apparent barrier $F^\ddagger$ for $\alpha$-relaxation. This can be done indirectly, by measuring the apparent activation barrier above and below the glass transition, respectively and then applying the the formalism presented in Ref.~\cite{LW_aging-sup}. For the sake of a quick estimate, one may use the available fragility data for the selenide from Ref.~\cite{https://doi.org/10.1111/jace.19491-sup}, whose authors report a fragility coefficient $m = 43$ at $T_g = 442\,\mathrm{K}$. Adopting $(\partial \ln K/\ln T)=-1$, one obtains $\Delta E^\ddagger \gtrsim 0.04$~$\mathrm{eV}$ for the dynamical range $\log_{10} [\tau_{\alpha}(T_g)/\tau_{\alpha, 0}] = 16$, $\Delta E^\ddagger \gtrsim 0.08$~$\mathrm{eV}$ for $\log_{10} [\tau_{\alpha}(T_g)/\tau_{\alpha, 0}] = 10$. This is consistent with Eq.~(7) being a lower bound.

Moving on to \cref{Edjump3Supple,Edjump4Supple}, the Kauzmann temperature $T_K$ is a fiducial temperature at which the configurational entropy would vanish. It is determined in two steps~\cite{L_AP-sup}: First one must subtract the entropy of a frozen glass from that of the equilibrated melt. We do not know the entropy of the glass but expect it to be numerically close to that of the corresponding crystal. There is, however, some uncertainty stemming from the free energy minima in glasses being separated by finite barriers in finite dimensions. In physical terms, this means that the glass might have additional degrees of freedom, including in particular the boson peak and two-level systems~\cite{LW-sup, LW_BP-sup, Lrelics-sup, LW_RMP-sup}, that might meaningfully contribute to its entropy, even though the contribution is expected to be modest~\cite{EastwoodW-sup}. Once evaluated, the configurational entropy must be extrapolated below the glass transition temperature to determine the temperature $T_K$ where it would vanish, which introduces further uncertainty. While the overall trend for the values of $T_K$ predicted using calorimetry and kinetics, respectively, clearly points out the two quantities are equal, individual substances may deviate from this trend substantially~\cite{RichertAngell-sup, L_AP-sup, RWLbarrier-sup, LRactivated-sup}.

\section{Conductivity prefactor}

Here we perform basic estimates for polaron-based mechanisms of electrical conduction, in which transport occurs via activated hops. First, we reproduce the expression from the main text for the conductivity in the present scenario 
\begin{equation} \label{Eq:cond2-sup}
  \sigma = \frac{q^2_e}{2 \pi \hbar \, \xi} \, \frac{\hbar \omega_B}{k_B T} \, e^{-\beta (E_v-\mu)}
\end{equation}
where for concreteness we assume that the hole polarons are the dominant charge carriers~\cite{Emin_rev-sup, Emin_revII-sup}. The insulating gap depends on the temperature, the dependence well approximated by a linear law within the temperature range in question~\cite{LKgap-sup}:
\begin{equation}
    E_g \approx \text{const} - \gamma T.
\end{equation}
According to our earlier work~\cite{LKgap-sup}, both edges of the insulating gap contribute to the temperature dependence of the insulating gap. The contribution of the temperature-dependent part of the gap to the prefactor in the conductivity is bounded from above by the quantity $e^{\gamma/2}$. Thus we proceed to estimate
\begin{equation}
    \text{prefactor} \lesssim \frac{q^2_e}{2 \pi \hbar \, \xi} \, \frac{\hbar \omega_B}{k_B T} \, e^{\gamma/2}
\end{equation}
For the four substances considered in Ref.~\cite{LKgap-sup}, the quantity $e^{\gamma/2}$ varies within the range $10^2$ to $10^3$. We adopt the lower value $10^2$, since the r.h.s. of the equation above is an upper bound. Near the glass transition, $\xi \simeq 2 \ldots 3$~nm~\cite{ZL_JCP-sup, LW-sup, L_AP-sup}.  The dimensionless quantity $\hbar \omega_B/k_B T$ is of order one. The effective charge $q_e$ of the polaron is numerically close to the electron charge, but is actually less, owing to the polarization of the lattice~\cite{LL2-sup}. Having this reduction in the effective charge in mind and setting $\xi = 3$~nm yields a prefactor of order $10^4$ or less, in agreement with Fig.~1 of the main text. 

If aging causes a relaxation in the electronic structure, there will be an additional contribution to $\Delta E^\ddagger$ for aged samples. Assume, again, that the dominant charge carrier is holes. It is conceivable that the Urbach tail states, which are candidate states for polarons~\cite{LL2-sup}, will relax relative to the edge of the valence band. We anticipate that this contribution would be much smaller than the present predictions for $\Delta E^\ddagger$.  Indeed, stabilization of electronic levels would have to be a small quotient of the enthalpy of melting per electron, which is roughly $1.5 k_B T_m/5$ or so for chalcogenides; here $T_m$ is the melting temperature. Indeed, the entropy of melting is rather universally $1.5 k_B$ per atom~\cite{CRC-sup} and the number of valence electrons, on the average, is close to five. In contrast, the predicted value of $\Delta E^\ddagger$ can be no smaller than $k_B T_g$ even for the strongest known substances. These notions are consistent with a reported insensitivity of the optical edge to heat treatment~\cite{ARAI1975295-sup}. 

It is instructive to compare the prefactor in Eq.~3 of main text to that arising in Emin's small polaron picture~\cite{Emin_rev-sup}, viz., $q_e^{2}\omega/(2 \pi l k_B T)$. Here, $l$ is the volumetric size of an atom, up to a constant of order one, and $\omega$ is a generic bond-vibrational frequency. Assuming, as before, that local vibrations are not sensitive to freezing, the apparent activation energy will exhibit a jump in its $T$-dependence, across the glass transition, because the lattice constant will increase with temperature at different rates below and above the glass transition, respectively. One may show straightforwardly that the resulting slope discontinuity is 
\begin{equation}
   \Delta E^\ddagger_V = \frac{k_B T_g^2}{3} \left[\frac{1}{V}\left(\frac{\partial V}{\partial T}\right)_{T_g^+} - \frac{1}{V}\left(\frac{\partial V}{\partial T}\right)_{T_g^-}\right],  
\end{equation}
 where $V^{-1}(\partial V/\partial T)$ is the thermal expansion coefficient. This yields $3.7\cdot 10^{-4}\,\mathrm{eV}$ for As$_2$Se$_{3}$, for which $V^{-1}(\partial V/\partial T)_{T_g^+} \approx 193\cdot 10^{-6}\,\mathrm{K}^{-1}$ and $V^{-1}(\partial V/\partial T)_{T_g^-} \approx 63\cdot 10^{-6}\,\mathrm{K}^{-1}$,  respectively~\cite{Voronova2001-sup}. We see the volumetric effect is two orders of magnitude less than the effect of the arrest of dynamics, below the glass transition, because the density depends on temperature too weakly in the first place.
In Emin's small polaron scenario, the prefactor would be increased, relative to the estimate above, by the factor $\xi/a \simeq 6$ or so. This is not a significant difference, in view of the approximate nature of the estimate in the first place.

%

\end{document}